\documentclass[aps,twocolumn,tightenlines,amsmath,amssymb,superscriptaddress]{revtex4-2}
\usepackage{graphicx}
\usepackage{xcolor}
\usepackage{color}
\usepackage[colorlinks=true]{hyperref}
\hypersetup{citecolor=black, urlcolor=black, linkcolor=black}
\usepackage{ulem}
\graphicspath{ {./images/} }
\usepackage{braket}
\usepackage{siunitx} 
\usepackage{orcidlink}
\usepackage{gensymb}
\begin{document}
\title{Realization of waveguide many-body quantum optics}
\author{Lena~M.~Hansen\,\orcidlink{0009-0003-9352-6246}}
\thanks{These authors contributed equally to this work.}
\affiliation{University of Vienna, Faculty of Physics, Vienna Center for Quantum Science and Technology (VCQ), 1090 Vienna, Austria}
\affiliation{Christian Doppler Laboratory for Photonic Quantum Computer, Faculty of Physics, University of Vienna, Vienna, Austria}
\author{Clara~Henke\,\orcidlink{0009-0004-5255-5689}} 
\thanks{These authors contributed equally to this work.}
\affiliation{Center for Hybrid Quantum Networks (Hy-Q), Niels Bohr Institute,
University of Copenhagen, Jagtvej 155A, Copenhagen DK-2200, Denmark}
\author{Christoph~Hotter\,\orcidlink{0009-0003-3854-0264}}
\affiliation{Center for Hybrid Quantum Networks (Hy-Q), Niels Bohr Institute,
University of Copenhagen, Jagtvej 155A, Copenhagen DK-2200, Denmark}
\affiliation{Department of Physics, MIT-Harvard Center for Ultracold Atoms and Research Laboratory of Electronics, Massachusetts Institute of Technology, Cambridge, Massachusetts 02139, USA}
\author{Oliver~A.~D.~Sandberg\,\orcidlink{0000-0002-9410-6150}}
\affiliation{Center for Hybrid Quantum Networks (Hy-Q), Niels Bohr Institute,
University of Copenhagen, Jagtvej 155A, Copenhagen DK-2200, Denmark}
\author{Thomas~Wilkens~Sandø\,\orcidlink{0009-0008-5931-2910}}
\affiliation{Center for Hybrid Quantum Networks (Hy-Q), Niels Bohr Institute,
University of Copenhagen, Jagtvej 155A, Copenhagen DK-2200, Denmark}
\author{Vasiliki~Angelopoulou\,\orcidlink{0000-0002-8115-3169}}
\affiliation{Center for Hybrid Quantum Networks (Hy-Q), Niels Bohr Institute,
University of Copenhagen, Jagtvej 155A, Copenhagen DK-2200, Denmark}
\author{Alexey~Tiranov\,\orcidlink{0000-0003-0791-8730}}
\affiliation{Center for Hybrid Quantum Networks (Hy-Q), Niels Bohr Institute,
University of Copenhagen, Jagtvej 155A, Copenhagen DK-2200, Denmark}
\author{Christoffer~B.~Møller\,\orcidlink{0000-0002-3123-0904}}
\affiliation{Center for Hybrid Quantum Networks (Hy-Q), Niels Bohr Institute,
University of Copenhagen, Jagtvej 155A, Copenhagen DK-2200, Denmark}
\author{Zhe~Liu\,\orcidlink{0000-0003-0672-4328}}
\affiliation{Center for Hybrid Quantum Networks (Hy-Q), Niels Bohr Institute,
University of Copenhagen, Jagtvej 155A, Copenhagen DK-2200, Denmark}
\author{Leonardo~Midolo\,\orcidlink{0000-0003-0237-587X}}
\affiliation{Center for Hybrid Quantum Networks (Hy-Q), Niels Bohr Institute,
University of Copenhagen, Jagtvej 155A, Copenhagen DK-2200, Denmark}
\author{Nikolai~Bart}
\affiliation{Lehrstuhl für Angewandte Festkörperphysik, Ruhr-Universität Bochum, Universitätsstraße 150, 44780 Bochum, Germany}
\author{Arne~Ludwig\,\orcidlink{0000-0002-2871-7789}}
\affiliation{Lehrstuhl für Angewandte Festkörperphysik, Ruhr-Universität Bochum, Universitätsstraße 150, 44780 Bochum, Germany}
\author{Philip~Walther\,\orcidlink{0000-0002-4964-817X}}
\affiliation{University of Vienna, Faculty of Physics, Vienna Center for Quantum Science and Technology (VCQ), 1090 Vienna, Austria}
\affiliation{Christian Doppler Laboratory for Photonic Quantum Computer, Faculty of Physics, University of Vienna, Vienna, Austria}
\author{Cornelis~J.~van~Diepen\,\orcidlink{0000-0001-8454-2859}}
\author{Peter~Lodahl\,\orcidlink{0000-0002-9348-9591}}
\affiliation{Center for Hybrid Quantum Networks (Hy-Q), Niels Bohr Institute,
University of Copenhagen, Jagtvej 155A, Copenhagen DK-2200, Denmark}
\author{Anders~Søndberg~Sørensen\,\orcidlink{0000-0003-1337-9163}}
\email[Corresponding author. Email: ]{anders.sorensen@nbi.ku.dk}
\affiliation{Center for Hybrid Quantum Networks (Hy-Q), Niels Bohr Institute,
University of Copenhagen, Jagtvej 155A, Copenhagen DK-2200, Denmark}
\begin{abstract}{
Controlling light photon-by-photon is central to quantum optics. 
At a fundamental level, photon interactions are mediated by their coupling to atoms, and ultimate control requires deterministic light-matter interfacing of single photons to single atoms. 
Extending this paradigm to radiatively couple multiple individual atoms in a deterministic and scalable manner opens the arena of many-body quantum optics. 
Here, we realize such a setting by coherently coupling solid-state artificial atoms to a nanophotonic waveguide and demonstrate higher-order photon correlations that are controlled by the number of quantum emitters. 
We study the scaling of nonlinear photonic transport induced by emitter-photon scattering and demonstrate that adding a quantum emitter generates higher-order photon correlations. 
Specifically, we experimentally observe genuine three-photon correlations from a pair of collectively coupled emitters, while contributions from lower photon numbers are suppressed. 
In addition, we scale to three resonant quantum emitters coupled to the waveguide. 
These advancements demonstrate the onset of many-body quantum optics in waveguide quantum electrodynamics, enabling new photonic quantum simulators, the creation of many-body entangled states, and the exploration of novel quantum phase transitions. 
}
\end{abstract}

\maketitle

Nonlinear interaction at the single-photon level is the fundamental principle underpinning the formation of many-body quantum correlations of light~\cite{Chang.2014}. 
The nonlinearity is ideally induced by a deterministic photon-atom interface, in which an atom mediates interactions between photons. 
Proof-of-concept photon-photon nonlinear interactions have been realized in both atomic and solid-state systems, including the demonstration of photon blockade~\cite{Birnbaum.2005}, attractive two-photon transport~\cite{Firstenberg.2013}, photon–photon quantum gates~\cite{Hacker.2016}, third-order correlations~\cite{Liang.16.2.2018, Stiesdal.2018, OrnelasHuerta.2021, Drori.13.7.2023}, bound-state dynamics~\cite{Tomm.2023}, and correlated and entangled two-photon states~\cite{LeJeannic.2022, Masters.2023, Liu.2024}. 
Extending quantum nonlinear optics to higher photon numbers and multiple emitters opens up new avenues for generating strong many-body correlations~\cite{Yudson.2008, Mahmoodian.2020, Prasad.2020}, non-Gaussian quantum optics~\cite{Walschaers.2021}, and simulating many-body effects~\cite{Noh.2017, Shi.2018, Fitzpatrick.2017, Fayard.2021, Kim.2021}. 
Recently, experiments have begun exploring this domain using ordered sub-wavelength arrays of emitters~\cite{Douglas.2026}. 
Accessing the realm of many-body quantum dynamics requires strong, coherent multi-photon interactions for a well-defined number of particles. 
Here, we utilize highly efficient light-matter waveguide interactions with a controllable number of atoms and exploit the resulting strong collective effects to implement quantum nonlinear optics in the multi-photon regime.

\begin{figure*}[htp!]
    \centering
    \includegraphics[width=0.75\textwidth]{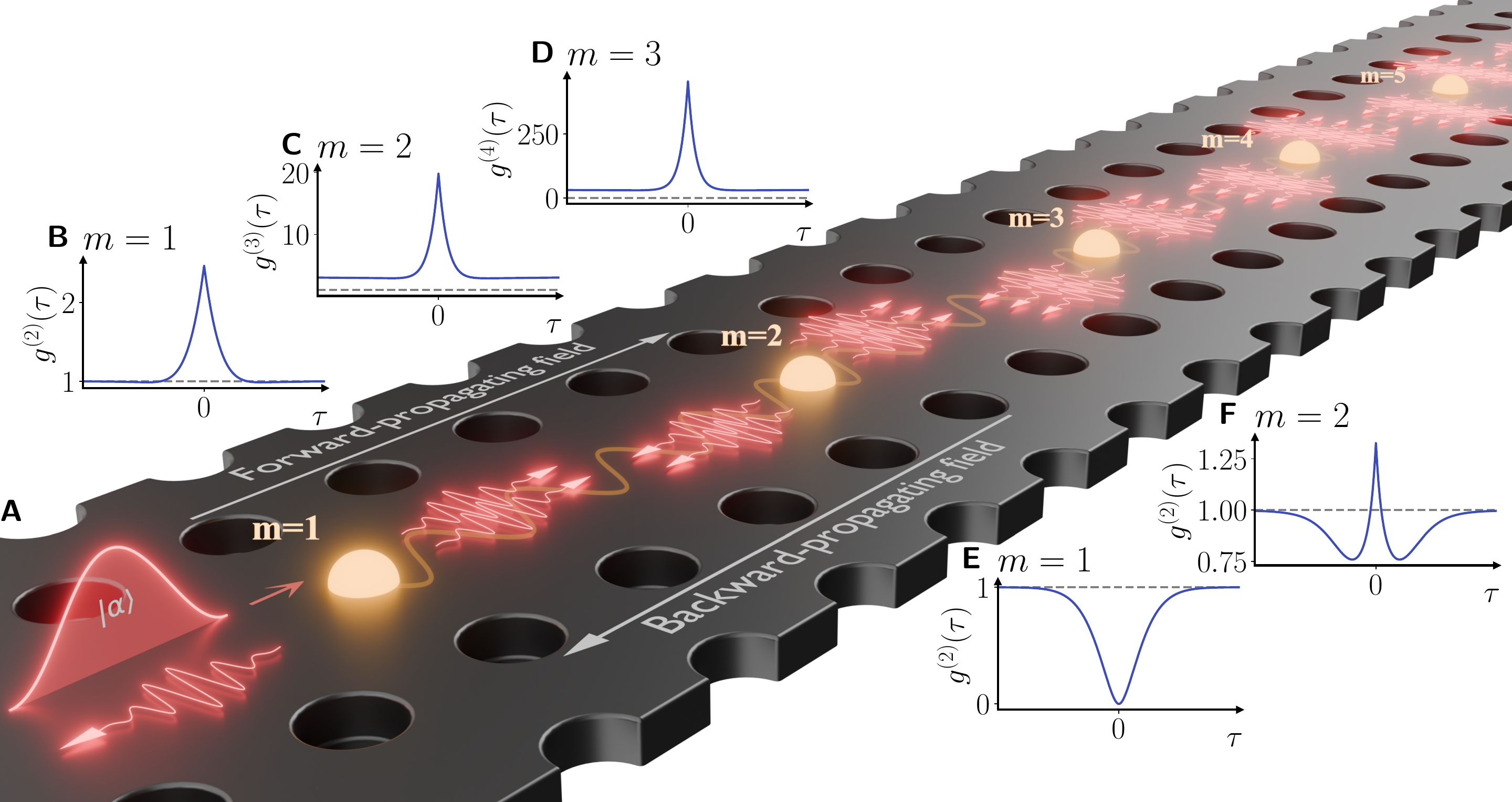}
    \caption{\textbf{Waveguide many-body quantum optics.} 
    (\textbf{A}) Illustration of multiple emitters directly coupled via the waveguide, visualized by a connection between the emitters. 
    The scattering of a weak coherent input pulse \(\ket{\alpha}\) in the bidirectional waveguide introduces a forward- and a backward-propagating field. 
    The photon wavepackets illustrate the underlying concept that \(m\) coupled emitters can tailor a forward-propagating output field with enhanced \(m+1\)-photon correlations and \(m\)-photon correlations in the backward direction. 
    (\textbf{B}-\textbf{D}) Simulations of the two-time correlation function $g^{(n)}(\tau)=\langle (a^\dagger(t))^{n-1}a^\dagger(t+\tau) a(t+\tau) (a(t))^{n-1} \rangle / \langle (a^\dagger(t)a(t) \rangle^n$ of the forward-propagating field. 
    In the forward-propagating field, the (\(m+1\))-order correlations are enhanced due to the interaction with \(m\) coupled emitters. 
    As $m$ increases, the photons become increasingly localized in time, corresponding to a superradiant burst.
    (\textbf{E} and \textbf{F}) Simulations of \(g^{(2)}(\tau)\) for \(m=1\) and \(m=2\) signaling the preference to backward scatter \(m\)th-order photon contributions.
    } 
\label{fig:0}
\end{figure*}

Engineered nanophotonic waveguides with solid-state emitters~\cite{Lodahl.2015, Sheremet.2023} offer a scalable route to realizing quantum nonlinear optics. 
Notably, the nonlinear light-matter interaction for a single emitter can be near deterministic due to the near-unity coupling efficiency~\cite{Arcari.2014}. 
Additionally, a waveguide geometry facilitates engineered, long-range dipole–dipole interactions between emitters~\cite{Tiranov.2023}. 
Thereby, strong emitter-photon and emitter-emitter coupling can be realized enabling strongly correlated photon transport which can be unraveled by photon-correlation measurements, see ~Fig.~\ref{fig:0}A. 
As a specific example, we show that strongly correlated states of \(m+1\) photons emerge from the interaction with \(m\) coupled emitters, see Fig.~\ref{fig:0}B-D. In addition, the reflection of the \(m\) coupled emitters reveals a preference for \(m\) photon scattering, as shown in Fig.~\ref{fig:0}, E and F. 
In contrast to sequentially generating highly complex photonic states through repeated interaction with a single emitter~\cite{Ferreira.2024}, the collective effects of coupled atomic systems open pathways for the direct formation of strongly correlated states through the simultaneous interaction of multiple photons and emitters.

In the present work, we realize the regime of many-body quantum optics by leveraging the quantum nonlinearity arising from collective effects of waveguide-coupled emitters. 
We report the observation of enhanced photonic correlations resulting from light-matter interactions with independently tunable quantum dots (QDs) embedded in a photonic crystal waveguide (PCW). 
By controlling two coupled emitters, we demonstrate strongly enhanced correlated three-photon states, while the incident one- and two-photon components are predominantly reflected.
These genuine three-photon correlations originate from the saturation of the fully excited emitter pair in conjunction with stimulated emission by the third photon. The result is a temporal burst of strongly correlated photons --- a process akin to superradiance~\cite{Dicke.1954}.
Additionally, by coupling a third emitter to the waveguide, we showcase the scalability of our approach, observing enhanced photon scattering with each added emitter. 
These results mark the onset of waveguide many-body quantum optics, where the interaction between multiple emitters and photons is controlled at the level of individual particles.
\subsection*{Three-photon dynamics from two coupled emitters}
We experimentally implement many-body quantum optics by scattering coherent light pulses off a pair of QDs coupled via a PCW. 
The device is two-sided, allowing for separate measurements of the forward- and backward-propagating photonic fields (for a detailed description of the device, see supplementary text, section~\ref{sec:qd_device}). 
Each emitter is individually controlled using electrostatic tuning via the Stark effect. 
When tuned into resonance, the emitters form a collective emitter system, mediated by the coupling to the waveguide mode~\cite{Tiranov.2023}. 
The coupling regime is determined by a phase lag of \(\phi = 0.8\pi\) between the emitters~\cite{Henke.2026}, resulting in an intermediate regime between dispersive and dissipative coupling. 
The coupling extends a distance of \(r=26\lambda\), where \(\lambda\) is the wavelength in the PCW. 
This remarkable long-range coupling is enabled by the waveguide, effectively turning the typical $1/r^3$ dipole-dipole interaction in a three-dimensional homogeneous environment to an infinite-range interaction in a one-dimensional waveguide geometry. 
This advancement is a crucial step for scalability, as it enables high-quality coupling of multiple emitters to the same optical mode, which is key to realizing many-body quantum optics. 
\begin{figure*}[htp!]
    \centering
    \includegraphics[width=1.0\textwidth]{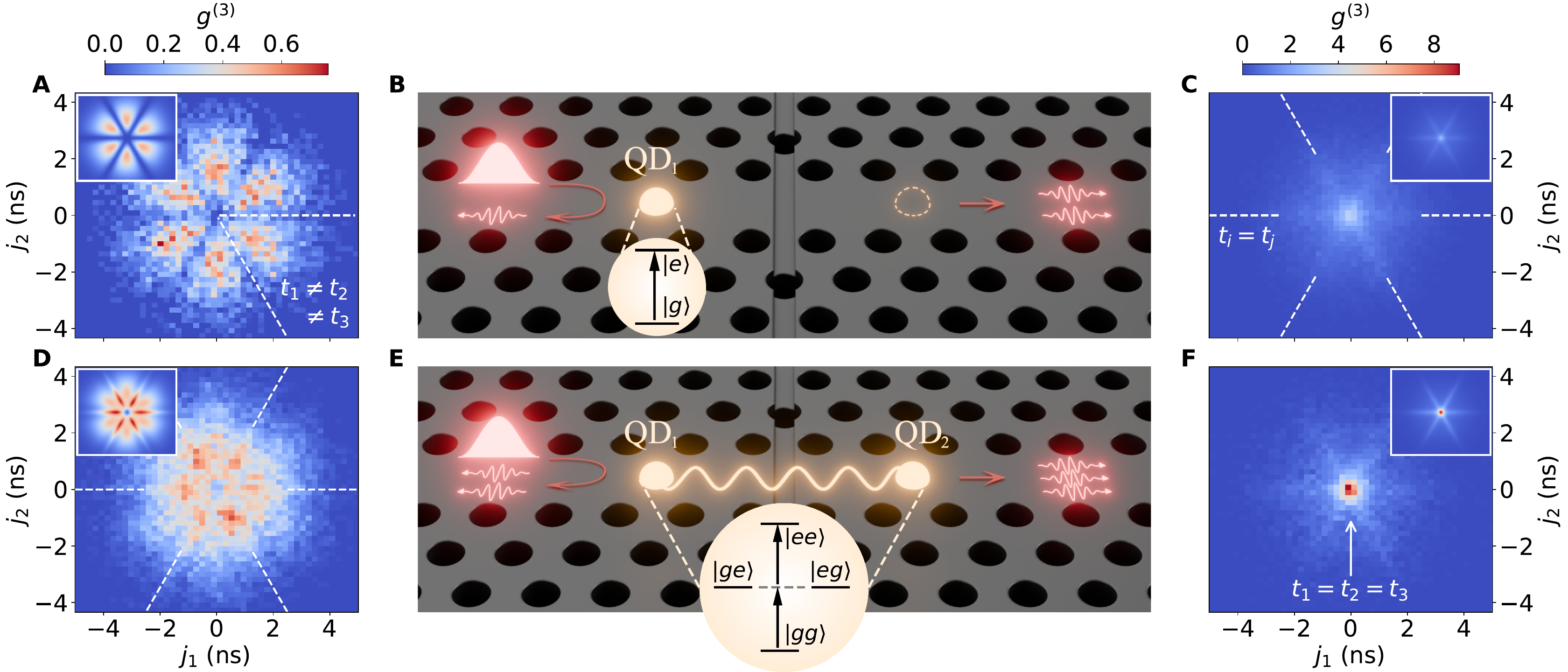}
    \caption{\textbf{Observation of strong three-photon correlations.} 
    (\textbf{A}-\textbf{C}) Waveguide scattering of a weak coherent light pulse for a single QD and (\textbf{D}-\textbf{F}) coupled QDs. 
    (\textbf{B} and \textbf{E}) Schematic of the device, where two quantum emitters ($\mathrm{QD}_1$ and $\mathrm{QD}_2$) are electrically separated by a vertical trench indicated in the center of the device, which allows independent electrical tuning of the QDs. The scattering process in the waveguide introduces forward- and backward-propagating fields (visualized by arrows). For the interaction with a single QD; a two-level system with ground state \(|g\rangle\) and excited state \(|e\rangle\), the second emitter is far-detuned from resonance (dashed line). The coupled emitter system is controlled by tuning both QDs into resonance. The coupled system comprises the collective ground state \(|gg\rangle\), the single-excitation states \(|ge\rangle\) and \(|eg\rangle\), and the collective excited state \(|ee\rangle\) via the coupling to the waveguide. 
    (\textbf{A} and \textbf{C}) The measured third-order correlations $g^{(3)}(t_1,t_2,t_3)$ are expressed in the Jacobi coordinates $j_1 = (2t_1 - t_2 - t_3)/\sqrt{6}$ and $j_2 = (t_2-t_3)/\sqrt{2}$ and summed over the center-of-mass coordinate $j_0 = (t_1 + t_2 + t_3)/\sqrt{3}$ resulting in an apparent sixfold rotational symmetry due to the long symmetric input pulse. 
    In the backward direction, a single QD scatters anti-correlated photons with \(t_1 \neq\!t_2 \neq t_3\). 
    In contrast, the forward direction shows correlations for $t_i = t_j$, where \(i, j\) are pairwise combinations of \(1, 2, 3\). 
    (\textbf{D}) The coupled QDs introduce detection events for $t_i = t_j$ in reflection but not for $t_1=t_2=t_3$, while (\textbf{F}) enhancing correlations at the origin $j_1 = j_2 = 0$ (corresponding to $t_1=t_2=t_3$) in the forward-propagating field. 
    The measurements for single QD and two coupled QDs are plotted with shared color bars for each direction. 
    The inset of each figure displays the corresponding theory simulation.} \vspace{-4pt}
\label{fig:1}
\end{figure*}

To investigate the multi-photon dynamics, we perform two scattering experiments, enabled by the tunability of the device, one with a single QD and the other with a pair of coupled QDs. 
Identical measurement conditions allow for a direct comparison between the two cases. 
In the single QD case (reference case), the transition of one QD is tuned to be resonant with the input light field, while the transition of the second QD is far detuned, yielding a two-level system (see Fig.~\ref{fig:1}B). 
For the case of coupled QDs, both QDs are tuned to be resonant with the input light field, leading to an effective multi-level system~\cite{Dicke.1954}. 
This allows us to investigate the collective nonlinear quantum effects of the coupled QDs (Fig.~\ref{fig:1}E). 

To study the many-body light-matter interaction, we extract $n$-photon correlation functions, $g^{(n)}(t_1, \ldots, t_n)$, with $t_i$ the detection time of photon $i$. In the experiment, the scattered field is split equally into three detection channels, enabling access to temporal correlations up to the third order (for a detailed description of the experimental setup, see supplementary text, section~\ref{sec:exp_setup}). 
The experiments are conducted with quasi-continuous pulses (see supplementary text, section~\ref{sec:excitation}), and in the weak-driving regime, with a mean photon number \(\langle n \rangle \leq 0.1\), such that contributions beyond third order are negligible. For comparison between the configurations with a single QD and two coupled QDs, the measurements are normalized to the recorded mean photon number. 

An elegant description of many-body systems is obtained using Jacobi coordinates. These naturally incorporate the system's symmetries and emphasize the relative temporal photonic coordinates (see caption Fig.~\ref{fig:1} for their explicit definition). 
The transformed correlation function, $g^{(3)}(j_1,j_2)$, after integration over the center-of-mass $j_0$, symmetrically encodes the full three-photon correlations~\cite{Stiesdal.2018}. 
By employing this description, we can interpret the many-body character of the light field and draw conclusions about temporal correlations involving up to three photons. 
\begin{figure}[htp!]
    \centering
    \includegraphics[width=0.48\textwidth]{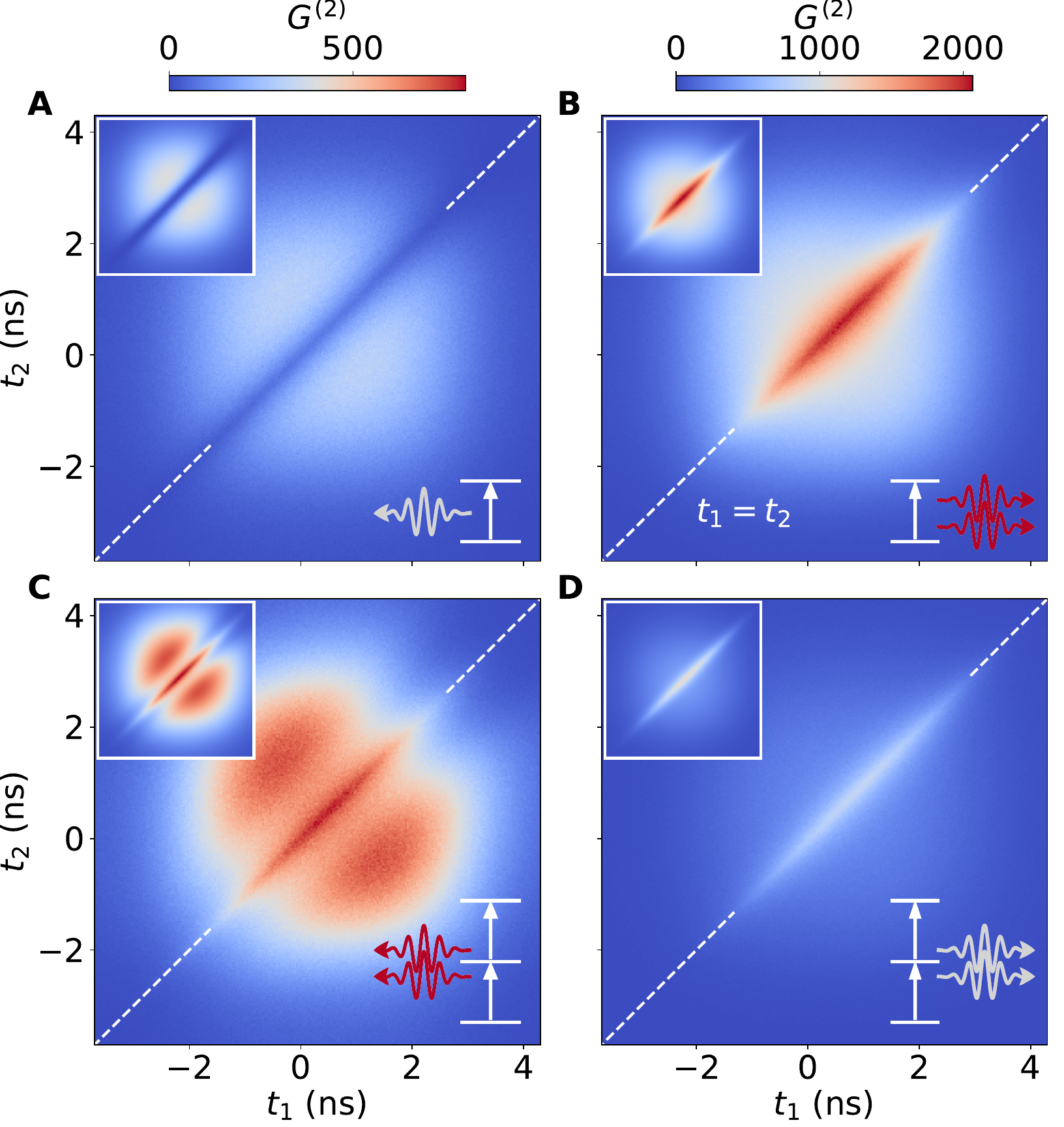}
    \caption{\textbf{Experimental two-photon correlations.} 
    Temporally resolved second-order correlations for (\textbf{A} and \textbf{B}) a single QD and (\textbf{C} and \textbf{D}) two coupled QDs. (\textbf{A}) The reflected field for a single QD consists only of anti-correlated photons, $t_1 \neq t_2$. (\textbf{C}) For coupled QDs, the reflection is generally increased, and in addition, a correlated component appears as a strong diagonal feature, $t_1 = t_2$. (\textbf{B}) For a single QD, the signal in the forward direction is dominated by a correlated two-photon component. (\textbf{D}) For coupled QDs, the signal in the forward direction is reduced, but a weak correlated component is still observed. 
    The coincidence counts for the QD and the QDs can be directly compared (shared colorbars) as they are performed for the identical measurement path, whereas the measurements for the forward and backward directions have different losses (separate colorbars). 
    The upper-left insets show theory simulations. Schematics in the lower right-hand corner indicate the energy levels, photon propagation direction, and dominant photon component, i.e., either one- or two-photon components. 
    The origin of the temporal coordinates \(t_1=t_2=0\) corresponds to the peak of the input pulse.
    } 
\label{fig:2}
\end{figure}

As a reference, we first consider the scattering from a single QD. 
The third-order correlations in the backward-propagating direction mimic a six-petal flower (Fig.~\ref{fig:1}A), while the forward-propagating field results in a six-armed star (Fig.~\ref{fig:1}C). 
Essentially, the backward direction resembles a negative image of the correlation map observed in the forward direction. 
The star comprises three symmetry axes with a relative angle of \(120 \degree\), which correspond to the pairwise combinations of time coordinates with $t_i = t_j$. 
These features reveal pairs of photons that are accompanied by an uncorrelated third photon, consistent with previous experiments on second-order correlations~\cite{Javadi.2015, LeJeannic.2022}. 
These second-order correlations rely on emitter-mediated photon-photon interactions, in which the single emitter can be excited by a first photon, and a second incident photon stimulates the emission~\cite{Hansen.30072024} in the forward direction, resulting in strong two-photon correlations. 
In contrast, the photon statistics in reflection consist only of coincidences with \(t_1 \neq t_2 \neq t_3\). 
This shows that solely single-photon components are present in the backward direction, since a single emitter can only reflect one photon at a time. 
The experimental observations agree well with the corresponding simulations (shown in the insets of each plot) and the predicted scattered photon wave function in reflection~\cite{Shen.2015}.

The photon scattering response of the collectively coupled QDs exhibits substantially modified third-order correlations. 
The experimental data (Fig.~\ref{fig:1}F) show a strong enhancement of the correlations at equal time $t_1\!=\!t_2\!=\!t_3$ ($j_1 = j_2 = 0$), compared to the single emitter case, in agreement with the theoretical simulations (shown in the inset). 
These higher-order correlations originate from the first two photons exciting the collectively coupled emitters, followed by a third photon stimulating a burst-like emission from the populated QDs. 
As a result, the zero-delay value of \(g^{(3)}_{\mathrm{QDs}}(0,0) = 8.4(5)\) well exceeds the single-emitter value of \(g^{(3)}_{\mathrm{QD}}(0,0) = 3.35(12)\). 
In the backward scattered field (Fig.~\ref{fig:1}D), the coupled emitters introduce two-photon correlations for \(t_i\!=\!t_j\), while still preserving an anti-bunched central feature. 
The prior is a result of populating the collective excited state by the absorption of two photons, and the subsequent emission of both photons in the backward direction. 
Two emitters, however, cannot simultaneously reflect three photons, resulting in the observed dip in the center at \(t_1=t_2=t_3\), which matches the strong three-photon correlations in the forward-propagating field of the coupled QDs.

To further dissect the nature of the observed third-order correlations, we note that they can contain both two-body and genuine three-body correlations. The former corresponds to correlated photon pairs plus an additional uncorrelated photon randomly located in time, while the latter originates from three-photon interactions. 
We investigate the two-body correlations by analyzing the time-resolved, unnormalized second-order correlations \(G^{(2)}(t_1, t_2)\) for each of the four cases discussed above (Fig.~\ref{fig:2}). 
In comparison to the third-order correlation data of Fig.~\ref{fig:1}, the second-order correlation data benefit from higher counting statistics, allowing for increased temporal resolution and a more accurate assessment of the single- and two-photon dynamics.

A single QD exhibits pronounced anti-bunching in reflection (Fig.~\ref{fig:2}A). 
The zero delay correlation value is $g^{(2)}_{\mathrm{QD}}(0)= 0.291(4)$, consistent with the emitter reflecting a single photon at a time. 
In the case of coupled QDs, a diagonal correlated feature with a superradiant decay appears for \(t_1=t_2\) in reflection (Fig.~\ref{fig:2}C). 
This enhanced diagonal, $g^{(2)}_{\mathrm{QDs}}(0)= 0.793(4)$, results from reflected photon pairs because the collectively excited state is resonant with two-photon processes. 
The correlated features for \(t_1=t_2\) are followed by decreased photon correlations, corresponding to a redistribution (bunching) of photons in time~\cite{vanDiepen.2025}. 
Notably, the single-photon reflection is also substantially enhanced for the pair of QDs, due to the increased scattering for multiple emitters~\cite{Chang.2012}.

\begin{figure*}[htp!]
    \centering
    \includegraphics[width=1.0\textwidth]{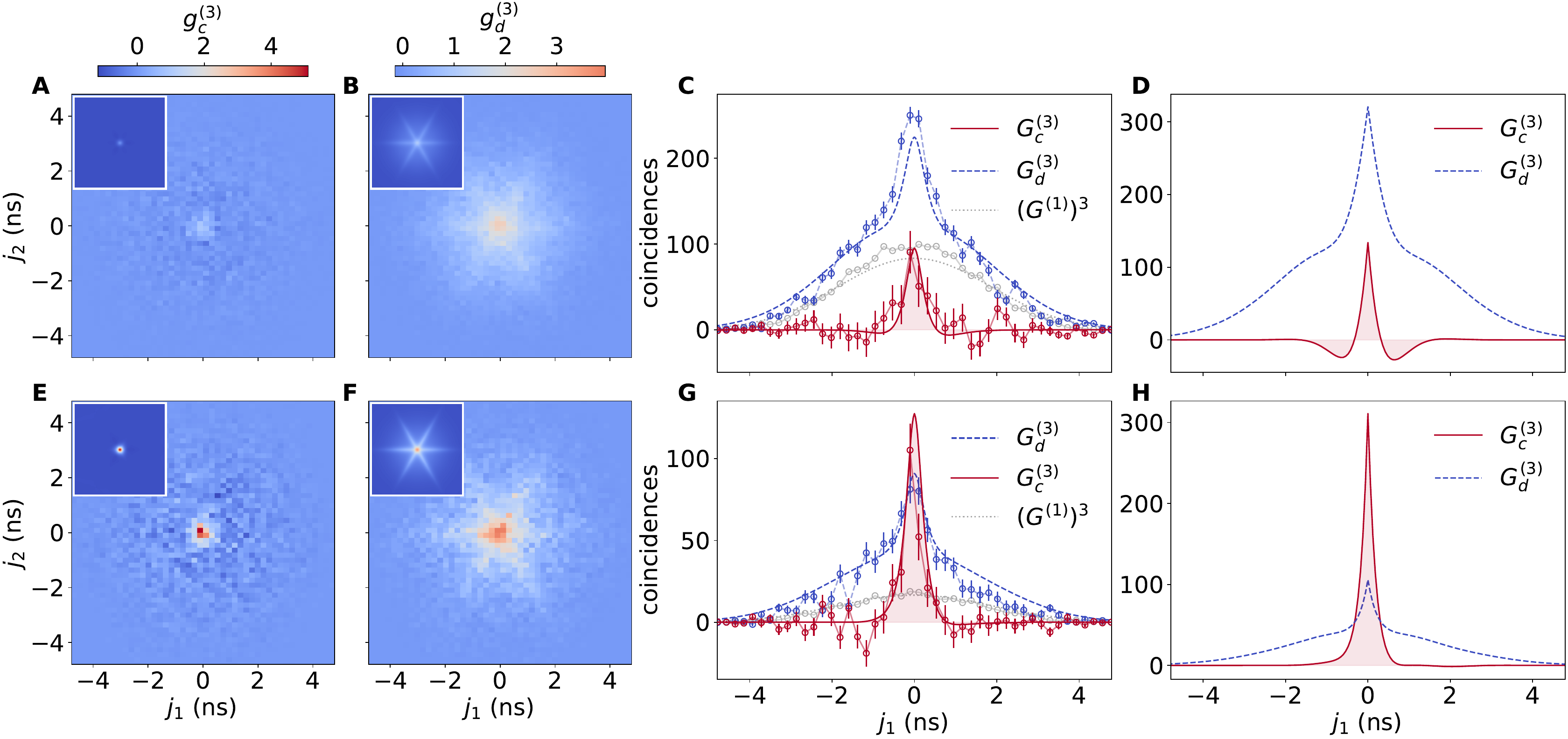}
    \caption{\textbf{Genuine three-photon correlations.} 
    Connected and disconnected components for the third-order correlations of the forward-propagating field for (\textbf{A}-\textbf{D}) a single QD and (\textbf{E}-\textbf{H}) a pair of coupled QDs. 
    Normalized (\textbf{A} and \textbf{E}) connected and (\textbf{B} and \textbf{F}) disconnected components. Insets show corresponding numerical simulations. 
    (\textbf{C} and \textbf{G}) Traces with \(j_2 = 0\) for the unnormalized correlations $G^{(3)}_c(j_1, j_2=0)$, and $G^{(3)}_d(j_1, j_2=0)$. 
    Additionally, the uncorrelated components \((G^{(1)})^3\) are shown as a reference. The overall amplitude of the theory simulations (lines) is fitted to the experimental data (dots), assuming uncorrelated spectral diffusion. The slight discrepancy in the overall amplitude fit arises from averaging the fitting amplitudes of both the single QD and the coupled QDs measurements. 
    The error bars indicate the uncertainties based on Poisson statistics and Gaussian error propagation. 
    (\textbf{D} and \textbf{H}) Simulations for $G^{(3)}_c(j_1, j_2=0)$ and $G^{(3)}_d(j_1, j_2=0)$ excluding spectral diffusion, and temporal jitter in detection.} 
\label{fig:3}
\end{figure*}

In the forward-propagating field, we observe strong two-photon correlations (\(t_1=t_2\)) for a single QD (Fig.~\ref{fig:2}B), resulting from the saturation of the emitter by a single photon followed by stimulated emission induced by a second photon. 
In comparison, this saturation effect, leading to the stimulated emission, is weaker for the coupled QDs (Fig.~\ref{fig:2}D), which can absorb two photons and decay superradiantly from the doubly excited state. The reemitted field interferes destructively with the transmitted field, a process that can be described as two-photon filtering, as it reduces the two-photon component. These results clearly demonstrate that the coupled system suppresses lower-order correlations in the forward-propagating field and instead they are directed to the backward-propagating mode.

\subsection*{Genuine three-photon correlations}
To accurately isolate the contribution of the genuine non-separable three-photon correlations from lower-order correlations, we analyze the third-order intensity fluctuations. 
This measure is also referred to as the third joint central moment, \(G^{(3)}_c = \langle :\delta n_1 \delta n_2 \delta n_3 :\rangle\), with \(\delta n_i = n_i - \langle n_i \rangle\), where $n_i$ is the number of photons going to detector $i$ and $\langle n_i \rangle$ its mean value. 
Up to third order, the joint central moment is equivalent to the joint cumulant~\cite{Kubo.1962, Stiesdal.2018, Plankensteiner.2022} and represents the connected component of the third-order correlation function, which contains only genuine three-photon correlations (see supplementary text, section~\ref{sec:3rd_connected} for theory on \(G^{(3)}_c\)). 
Accordingly, the contribution from the disconnected component -- which only involves lower-order correlations -- can be determined by \(G^{(3)}_d = G^{(3)} - G^{(3)}_c\)~\cite{Drori.13.7.2023}. 
Experimentally, we analyze the connected and disconnected components by recording fully correlated measurements within a single pulse duration, partially correlated measurements involving two consecutive pulses, and uncorrelated measurements from three consecutive pulses (for the complete data set, see supplementary text, section~\ref{sec:complete_dataset}). 
The normalized joint cumulant, \(g^{(3)}_c(j_1, j_2)\), is obtained by normalization with the central region (\(j_1=j_2=0\)) of the uncorrelated measurement. 
Genuine three-photon correlations thus result in a positive contribution at the origin \(t_1=t_2=t_3\). 

In Figure~\ref{fig:3}, we show the connected and disconnected three-photon components of the forward-propagating field. As a reference, an uncorrelated coherent input field is characterized by \(g^{(3)}_c(0,0) = 0\). For the single QD (Fig.~\ref{fig:3}A), a weak connected three-photon component of \(g^{(3)}_c(0, 0)=0.83(5)\) is measured, resulting primarily from removing single-photon components from the coherent input field (see supplementary text, section \ref{sec:3rd_connected}). 
Remarkably, for the coupled QDs (Fig.~\ref{fig:3}E), a strong enhancement of the genuine three-photon correlations is observed with a clear central feature of \(g^{(3)}_c(0, 0)= 4.3(2)\). These measurements are in agreement with the theory simulations, which are shown in the insets. The measured maximal value of the enhancement depends on the temporal resolution of the measurement. 
The corresponding simulations with(without) convolution of the experimental detection resolution yield \(g^{(3)}_c(0,0) = 5.4\) (\(g^{(3)}_c(0,0) = 7.7\)) for coupled QDs relative to \(g^{(3)}_c(0, 0) = 0.8\) (\(g^{(3)}_c(0,0) = 1.1\)) for a single QD (see supplementary text, section~\ref{sec:norm_avg}). In contrast to the pronounced central feature of the connected three-photon component \(g^{(3)}_c(j_1, j_2)\), the disconnected three-photon component $g^{(3)}_d(j_1, j_2)$ (Fig.~\ref{fig:3}, B and F) exhibits the six-armed star temporal structure resulting from the dominant correlated two-photon scattering.

The temporal profiles of the unnormalized correlations \(G^{(3)}_c(j_1, j_2=0)\), and \(G^{(3)}_d(j_1, j_2=0)\) (Fig.~\ref{fig:3}, C and G) illustrate the contribution of the disconnected component and the connected component to the third-order correlations. As a reference, the uncorrelated components from independent scattering, $(G^{(1)}(j_1, j_2=0))^3$, show only very limited temporal structure since they are dominated by independent single-photon scattering.
In the case of a single emitter, the most significant contribution is the disconnected component, which features pronounced side arms and a broad background of uncorrelated events during the pulse duration. 
On the contrary, for the coupled QDs, the disconnected features \(G^{(3)}_d\) are strongly suppressed in the third-order correlations. 
Instead, the output field consists of enhanced connected three-photon correlations in the center (\(j_1=j_2=0\)). 
This is clearly demonstrated by an increased difference in the temporal profiles between the reference of uncorrelated components, $(G^{(1)})^3$, and the connected three-photon correlations \(G^{(3)}_c\). 
These results demonstrate that the genuine three-photon correlations are the dominant three-photon contribution to the output field at \(j_1,j_2\approx 0\) for the collective coupled system. 
Simulations, shown in Fig.~\ref{fig:3}, D and H, indicate that the genuine three-photon correlations can be further enhanced by reducing spectral diffusion and detector jitter. 
Furthermore, we identify that the strongest collective quantum nonlinear effects can be explored in the fully dissipative or dispersive coupling regime, see supplementary text, section~\ref{sec:phase_scan}. 
Essentially, the genuine three-photon correlations are caused by the coupled system containing a doubly excited state but not a triply excited state. For a single QD, quantum nonlinearity sets in at the level of two photons, whereas strong nonlinearity mainly appears for three photons for the two coupled QDs.
The connected three-photon correlations are strongly correlated in time compared to other contributions, consistent with a superradiant burst from the collectively excited state stimulated by the third photon. 
\subsection*{Scaling to multiple emitters}

\begin{figure}[htp!]
    \centering
    \includegraphics[width=0.48\textwidth]{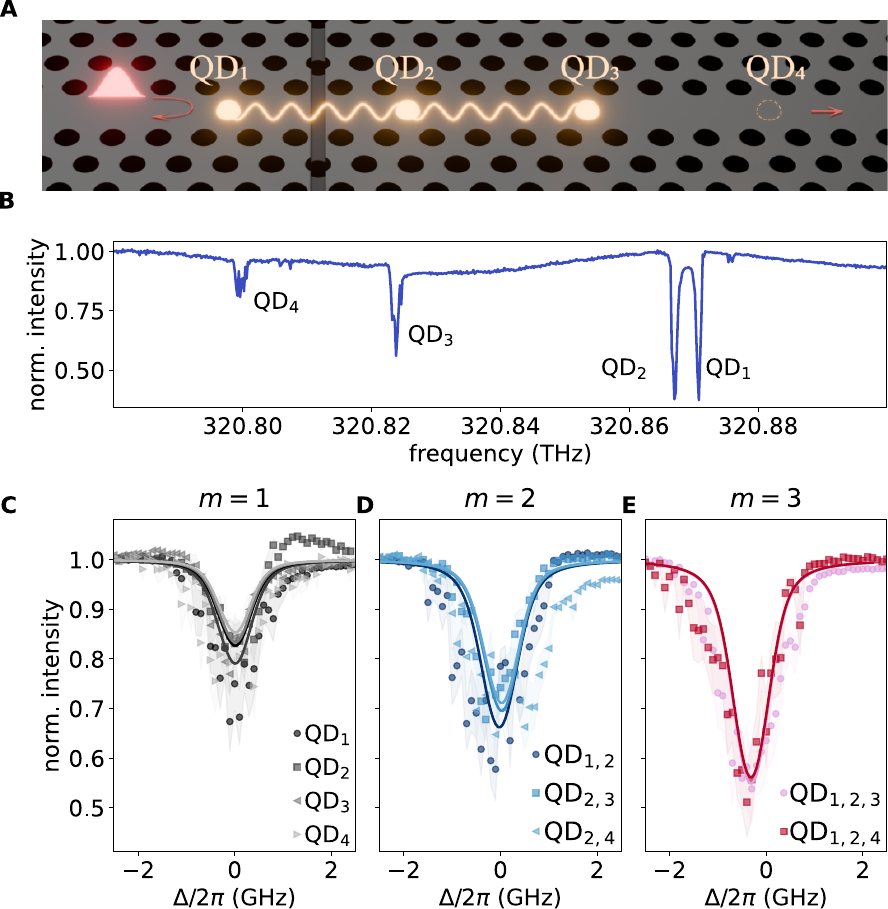}
    \caption{\textbf{Scattering from multiple emitters.} (\textbf{A}) Visualization of three coupled emitters formed from a set of four QDs in a waveguide. (\textbf{B}) Transmission spectrum of four emitters within the waveguide. (\textbf{C-E}) Experimental data (markers) and simulations (lines) of the transmission intensity for simultaneously tuning (\textbf{C}) single emitters, (\textbf{D}) pairs of emitters, and (\textbf{E}) triplets of emitters into resonance with the input field using electric and magnetic tuning. The transmission intensity is normalized to the far off-resonant detuning. The shaded areas indicate the standard error of the mean across 20 measurement runs. The theory plots for $m=3$ are shifted by $\SI{0.3}{GHz}$ to align with the experimental data.}
\label{fig:4}
\end{figure}

An important aspect of many-body waveguide quantum optics is the scaling to multiple emitters. Our device contains four spectrally close emitters, of which combinations of three emitters can be tuned into resonance, as illustrated in Fig.~\ref{fig:4}A. The transmission spectrum, shown in Fig.~\ref{fig:4}B, demonstrates distinct transmission dips of the four individual emitters. In addition to controlling up to two emitters independently using electrical control~\cite{Henke.2026}, a third emitter is brought into resonance through Zeeman tuning with a magnetic field~\cite{Tiranov.2023}. Detailed information is provided in the supplementary text, section~\ref{sec:3QDs}.

The collective effects of multiple emitters are investigated by measuring the intensity of the forward-propagating field through the waveguide for \(m=1,2,3\) emitters tuned into resonance. For two emitters (Fig.~\ref{fig:4}D), the measurements show that the intensity suppression is more pronounced than for a single emitter (Fig.~\ref{fig:4}C). When three emitters are on resonance (Fig.~\ref{fig:4}E), the suppression is further enhanced, leading to a deeper transmission dip. These results demonstrate that the increased reflection scales with the number of emitters \(m\) due to the enhanced collective light-matter interaction when more emitters are tuned into resonance. 
This behavior aligns with the simulation results of coupled QDs and is consistent with the coupling of three emitters within the waveguide. We note that the transmission spectrum of the three coupled QDs shows a frequency shift for the measurement, which is considerably smaller in the simulation. Therefore, the theory curve has been shifted to match the data. Such shifts may arise from the Fano profile of the line shapes caused by reflections at the waveguide ends, which is not included in the simulation. Alternatively, the shifts might be attributed to experimental calibration errors, or in fact be a signature of a collective Lamb shift in single-photon superradiance~\cite{Scully.2009}. 
The complete set of measurements and the corresponding theory simulations are included in the supplementary text, section~\ref{sec:3QDs}.

In addition to opening a new avenue for studies of multi-emitter arrays~\cite{Chang.2018}, the ability to tune multiple emitters into resonance allows engineering of higher-order photon correlations in scattering experiments. The universality of this phenomenon is explored in the supplementary text, section~\ref{sec:scaling}, where we investigate higher-order connected correlations for up to \(m = 5\) coupled QDs. We find that the highest value of connected correlations \(g^{(n)}_c\) in the forward-propagating field is achieved when \(n=m+1\). These results suggest a scalable method for controlling correlations on a photon-by-photon basis, mediated by the coupled multi-emitter system.

\subsection*{Conclusions}
We have reported on the experimental demonstration of strongly correlated three-photon dynamics using a pair of QDs due to the strong multi-photon interactions mediated by collective coupling through a waveguide. In essence, this effect arises from an elementary principle: two coupled QDs can absorb at most two photons simultaneously. Consequently, strong three-photon correlations emerge from the coupled system. The enhancement of genuine three-photon correlations corresponds to a superradiant burst from the collectively excited state, stimulated by a third photon.

We demonstrate an essential aspect for advancing many-body waveguide QED by controlling a three-emitter system coupled to the waveguide. Controlling multiple emitters strongly coupled to the light mode opens new avenues for many-body quantum optics, as we find that the enhancement of higher-order correlations can be generalized to larger numbers of coupled emitters. In prospective implementations, individual local tuning can be integrated for each emitter, enabling precise control of the correlated multi-photon state mediated by the emitter system.

The present experiment marks the advent of strongly correlated many-body quantum optics, featuring strong interactions between emitters and photons. Controlling the individual components of the system allows engineering non-equilibrium many-body systems, one emitter and one photon at a time. This approach enables new possibilities for realizing many-body physics of strongly correlated light-matter systems for investigating novel quantum phase transitions and photonic quantum simulation~\cite{Noh.2017}, many-body entangled states~\cite{Mahmoodian.2020, Gonzalez-Tudela.2015}, and quantum logic for photonic quantum computing~\cite{Dzsotjan.2010}. 

\bibliography{bib.bib}

\noindent{\bf Data availability}~Derived data supporting the findings of this study are available from L.M.H. upon request.

\noindent{\bf Code availability}~The numerical simulation codes used in this work are available from Ch.H. upon request.

\noindent{\bf Acknowledgments}~We are grateful to Klaus Mølmer, Ying Wang, and Noah Kaufmann for fruitful discussions.

\noindent{\bf Author contributions}~L.M.H., C.J.v.D., P.L., and A.S.S. conceived the project. C.H., L.M.H., and C.J.v.D. carried out the measurements. V.A., A.T., and C.J.v.D. upgraded the optical setup and designed the sample. Z.L. fabricated the device, and L.M. developed the fabrication process. N.B. and A.L. carried out the growth and design of the wafer. C.H., T.W.S., and C.J.v.D. characterized the sample. C.B.M. assisted with the measurements of three emitters. Ch.H., O.A.D.S., and A.S.S. developed the theoretical model. Ch.H. and O.A.D.S. carried out the theoretical simulations. L.M.H., C.H., and C.J.v.D. analyzed the data and prepared the figures. P.W. supervised the work carried out at the University of Vienna. L.M.H., C.H., Ch.H., C.J.v.D., P.L., and A.S.S. wrote the manuscript. All authors commented on the manuscript. C.J.v.D., P.L., and A.S.S. supervised the project.

\noindent{\bf Funding}~We gratefully acknowledge financial support from Danmarks Grundforskningsfond (DNRF 139, Hy-Q Center for Hybrid Quantum Networks), the Novo Nordisk Foundation (Challenge project “Solid-Q”), Novo Nordisk Foundation (NNF20OC0060019), Vienna Doctoral School in Physics (VDSP), BMFTR funded projects QTRAIN No. 13N17328, EQSOTIC No. 16KIS2061, and QR.N No.~16KIS2200, and the DFG funded project EXC ML4Q LU 2004/2 No.~390534769. The financial support by the Austrian Federal Ministry of Labour and Economy, the National Foundation for Research, Technology and Development and the Christian Doppler Research Association is gratefully acknowledged. Funded (Co-funded) by the European Union (HORIZON Europe Research and Innovation Programme, EPIQUE, No~101135288). Views and opinions expressed are however those of the author(s) only and do not necessarily reflect those of the European Union or the European Commission-EU. Neither the European Union nor the granting authority can be held responsible for them. 
This research was funded in whole or in part by the Austrian Science Fund (FWF)[10.55776/F71]. Ch.H. was supported by the Carlsberg Foundation through the ``Semper Ardens" Research Project QCooL. This research was funded in part by the Austrian Science Fund (FWF) 10.55776/J4865. For the purpose of open access, the author has applied a CC BY public copyright license to any Author Accepted Manuscript version arising from this submission.

\noindent{\bf Competing interests}~P.~L. is founder of the company Sparrow Quantum, which commercializes single-photon sources. The authors declare no other competing interests.

\noindent{\bf Supplementary Information} is available for this paper. 

\noindent{\bf Correspondence and requests for materials} should be addressed to Anders Søndberg Sørensen.

\newpage
\clearpage

\renewcommand{\thefigure}{S\arabic{figure}}
\setcounter{section}{0}
\renewcommand{\thesubsection}{\arabic{subsection}}
\makeatletter
\renewcommand{\p@subsection}{S}
\renewcommand{\thesubsubsection}{\arabic{subsubsection}}  
\renewcommand{\p@subsubsection}{S\thesubsection.} 
\makeatother
\onecolumngrid
\setcounter{figure}{0}
\section*{Supplementary Information}
\subsection{Quantum dot device}\label{sec:qd_device}

\begin{figure}[htp!]
    \centering
    \includegraphics[width=0.9\textwidth]{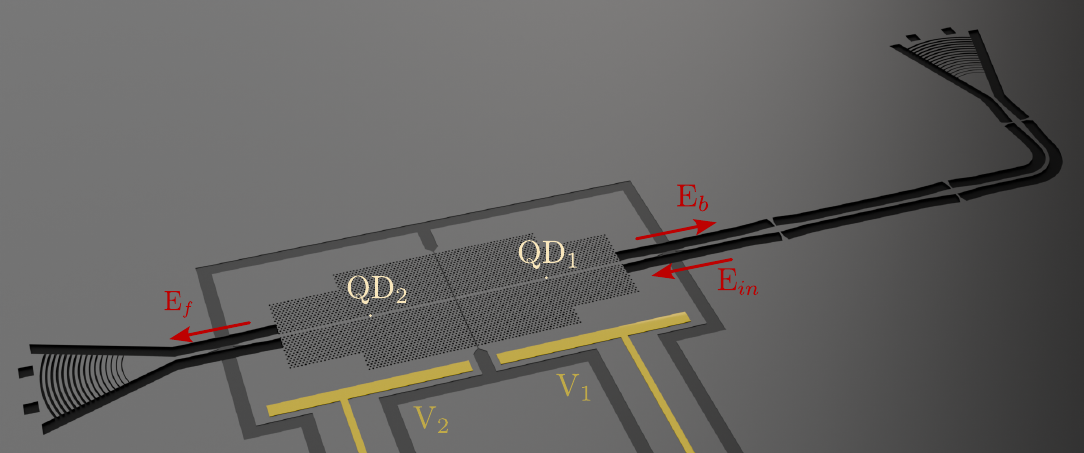}
    \caption{\textbf{Quantum dot device.} The graphic shows the structure of the QD device. Two QDs (QD$_1$ and QD$_2$) are coupled to a PCW. The trenches (marked in dark gray) electrically isolate each of the emitters, such that they can be independently controlled with local electric Stark fields by applying gate voltages ($V_1$ and $V_2$) at the contacts. For the scattering experiments, weak-coherent laser pulses ($E_{\mathrm{in}}$) are coupled from the right-hand grating coupler into the device. The forward- and backward propagating output fields ($E_{\mathrm{f}}$ and $E_{\mathrm{b}}$) are collected at the left- and right-hand grating couplers, respectively.} 
\label{fig:chip}
\end{figure}

InAs quantum dots (QDs) are embedded in a GaAs wafer that features a p-i-n diode heterostructure. Each nano-photonic device (see Fig.~\ref{fig:chip}) consists of a photonic crystal waveguide (PCW) with grating couplers connected at both ends via nanobeam waveguides. A shallow trench is etched into the center of the PCW to enable independent electrical control of the left and right halves. Further details regarding the wafer and waveguide fabrication are provided in Ref.~\cite{Uppu2020}.
On resonance, collective effects are induced by the dipole-dipole interaction between the QDs, which is mediated via the waveguide and depends on the exact position of each emitter with respect to each other and with respect to the unit cell of the PCW corresponding to a phase lag accumulated between them~\cite{Tiranov.2023}. 

The characterization of the QDs is described in detail in Ref.~\cite{Henke.2026}. For convenience, an overview of the extracted parameters is given in Table~\ref{table:1}. The effective coupling phase $\phi=0.8\pi$ between QD$_1$ and QD$_2$ indicates a coupling behavior that comprises a dispersive and dissipative component. Important to highlight is the long coupling range of \SI{13}{um} between QD$_1$ and QD$_2$, which is approximately 26$\lambda$, where $\lambda\approx \SI{500}{nm}$ is the wavelength in GaAs (free space wavelength $\approx \SI{940}{nm}$). The emitters are $\approx\SI{3}{THz}$ away from the band edge of the PCW (see Ref.~\cite{Henke.2026}).

\begin{table}[h!]
\centering
\begin{tabular}{c|c|c}
     & QD$_1$ & QD$_2$   \\
     \hline\hline
    $\Gamma/ 2 \pi$, GHz &0.388(2)& 0.345(2)\\ \hline
    $\gamma_d / 2 \pi$, GHz&0.01&0.09\\ \hline
    $\beta$ & 0.95&0.85  \\ \hline
    $\sigma_{sd} / 2\pi$, GHz & 0.30&0.22 \\ \hline
    $\xi/2\pi$, GHz &0.04& 0.19  \\ \hline
    $\sigma_{IRF}$, ps& \multicolumn{2}{c}{83} \\  \hline
    $\phi$ & \multicolumn{2}{c}{0.8 $\pi$} \\
\end{tabular}
\caption{\textbf{Parameters of the quantum dots.} Decay rate $\Gamma$, dephasing rate $\gamma_d$, $\beta$ factor, spectral diffusion $\sigma_{sd}$, Fano factor $\xi$, the jitter of the detection setup modeled as a Gaussian with standard deviation $\sigma_{IRF}$, and the coupling phase between the two QDs $\phi$.} 
\label{table:1}
\end{table}

\subsection{Experimental setup.}\label{sec:exp_setup}
\begin{figure}[htp!]
    \centering
    \includegraphics[width=0.75\textwidth]{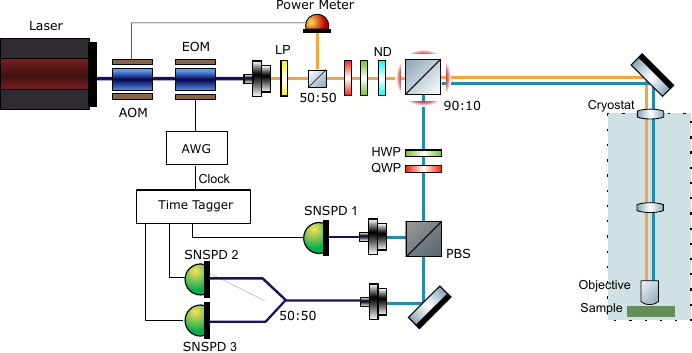}
    \caption{\textbf{Experimental setup} For the excitation (orange path), a CW laser is shaped using an EOM controlled by an AWG. The power of the pulses is controlled by an AOM and neutral-density (ND) filters. The polarization is controlled using a linear polarizer (LP), a quarter-wave plate (QWP), and a half-wave plate (HWP). The collected signal (blue path) is split into three paths using a PBS and a subsequent fiber beam splitter. Photons are detected on SNSPDs and their arrival times are recorded by a Time Tagger. }
    \label{fig:setup}
\end{figure}

A schematic of the setup is shown in Fig.~\ref{fig:setup}. The sample is placed in a cryostat at a temperature of \SI{4}{K}. The lens system in the cryostat comprises a 4f system to be more robust towards imperfect collimation and an objective to focus the light onto the device. For the excitation, a continuous-wave (CW) laser is shaped into Gaussian pulses using an electro-optic modulator (EOM) controlled by an arbitrary waveform generator (AWG). The power of the pulses is controlled by an acousto-optic modulator (AOM) and measured by a power meter. For the reflection measurements, we use a cross-polarization scheme to suppress input laser light in the collection. For the transmission measurements, we use spatial suppression to select between input and output path. The collected signal is split into three paths using a polarizing beam splitter (PBS) and a subsequent fiber beam splitter (BS). Changing the waveplates in front of the PBS allows to obtain a balanced output with equal counts at each superconducting nanowire single photon detector (SNSPD). A more detailed description of the excitation, collection, and cross-polarization is given in the subsections below.

\subsubsection{Excitation}\label{sec:excitation}
\begin{figure}[h!]
    \centering
    \includegraphics[width=0.4\textwidth]{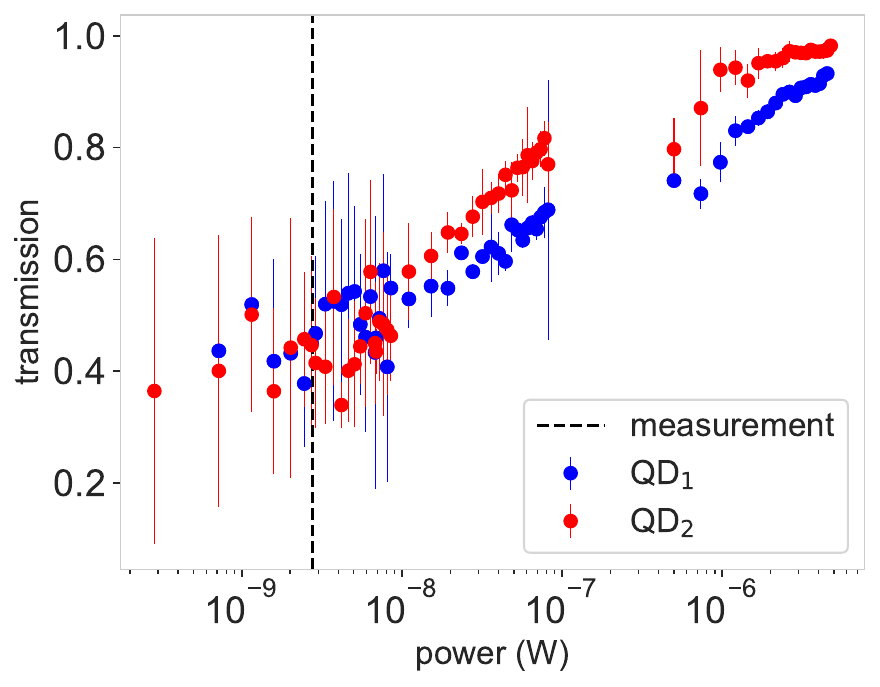}
    \caption{\textbf{Calibration of average photon number in a QD lifetime.} Minimum transmission intensity for different pump powers extracted from varying the detuning between QD and laser for QD$_1$ (blue) and QD$_2$ (red). Error bars represent the standard deviation between 10 different sets of measurements. The experiments in the main text are conducted below saturation with the average number of photons in a QD lifetime $\langle n\rangle\leq0.1$ (black dashed line).} 
\label{fig:RT_power}
\end{figure}

To generate weak-coherent pulses, a narrow-bandwidth CW laser (Toptica CTL) is temporally shaped using an EOM driven by an AWG. The laser frequency is set to \SI{320.8615}{THz} and the AWG generates pulses at a repetition rate of \SI{50}{MHz}. The input pulse width of \(\sigma = \SI{3}{ns}\) is quasi continuous relative to the radiative linewidths of the single emitters (QD$_1$ and QD$_2$) with \(\Gamma_1/2\pi = 0.388(2)\,\text{GHz}\) and \(\Gamma_2/2\pi = 0.345(2)\,\text{GHz}\), with \(\sigma\Gamma_1 = 7.3\) and \(\sigma\Gamma_2 = 6.5\).
\\
The effective laser power is attenuated below the saturation power of the QDs, while still remaining high enough to generate a sufficient number of coincidence events. The power is calibrated by a series of resonant transmission measurements for each QD with three different ND filters in the excitation path to cover a broader range of effective powers. The transmissivity of each ND filter is $\eta_{ND50}=0.028\,\%$, $\eta_{ND40}=0.27\,\%$, $\eta_{ND10}=10.49\,\%$. The power recorded before the ND filters is corrected for these values to obtain the effective power after. Figure~\ref{fig:RT_power} shows the minimum transmission dip for different powers, averaged over 10 measurement runs. The average photon number in a QD lifetime is $\langle n \rangle=\frac{\Omega^2}{2\beta\Gamma^2}$ with Rabi frequency $\Omega=2\sqrt{\eta P}$ with $\eta$ a proportionality factor that accounts for losses of the setup and the power~$P$. For the power used during the scattering experiments (dashed black line in Fig.~\ref{fig:RT_power}), we find $\langle n \rangle\leq 0.1$. A proportional–integral–derivative (PID) controller takes as input the power measured by the power meter and outputs a corresponding signal to the AOM to set and stabilize the laser power.

\subsubsection{Collection}
The collected light is routed into three detection channels using two cascaded beam splitters: a PBS, followed by a fiber-based 50:50 beam splitter placed in one of the PBS output arms. A waveplate preceding the PBS is adjusted to ensure equal count rates across the three channels. Since the detection efficiency of the SNSPDs is polarization dependent, fiber-based polarization control is used to maximize the count rate (omitted in Fig~\ref{fig:setup} for visual clarity). The detected events are correlated using a Time Tagger. Temporal offsets between detection paths are calibrated and corrected using the Time Tagger software. 
\subsubsection{Cross-polarization}
To suppress noise from reflections at the sample surface, the excitation and collection paths are cross-polarized using a HWP and a QWP in the respective path. As the light from the grating couplers is orthogonally polarized, the waveplates in the forward-scattering experiment are adjusted to match the polarizations of the input and output from the grating couplers. For the backward-scattering experiment, the excitation polarization is set \SI{45}{\degree} relative to the optimal input grating coupler polarization, while the collection path is aligned orthogonally. The excitation laser power is adjusted to maintain a constant effective power within the waveguide for both forward- and backward-scattering experiments.

\subsubsection{Quantum dot detuning} \label{sec:QD_detuning}
\begin{figure}[h!]
    \centering
    \includegraphics[width=0.7\linewidth]{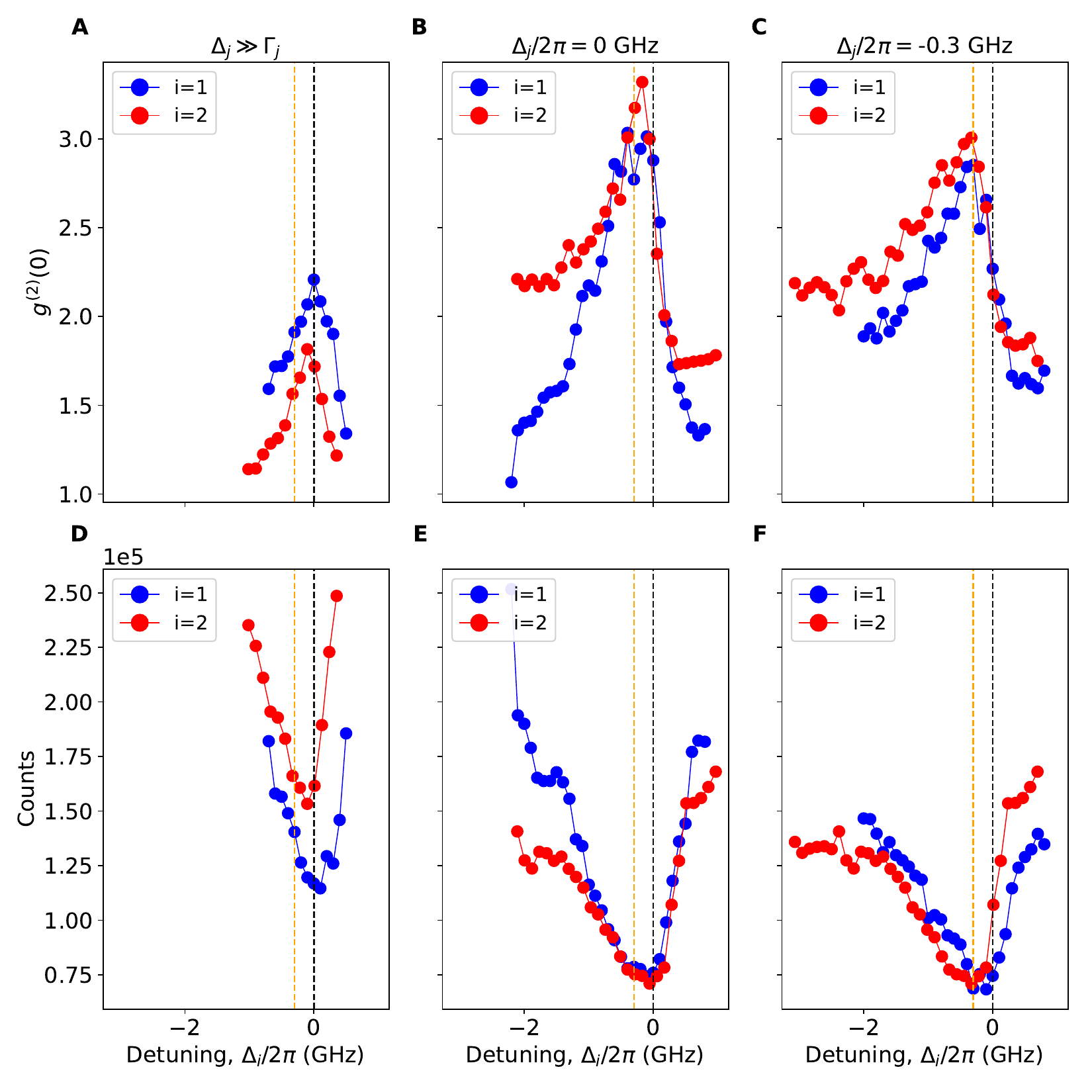}
    \caption{\textbf{Second-order correlation and intensity of forward propagating field for different detunings.} Measurement of second-order correlation $g^{(2)}(0)$ and intensity as a function of the detuning $\Delta_i$ between QD$_i$ and the resonant laser for QD$_1$ (i=1, blue) and QD$_2$ (i=2, red). When the other QD is far detuned, $\Delta_j/2\pi\gg\Gamma_j/2\pi$ (\textbf{A} and \textbf{D}), on resonance with the laser $\Delta_j/2\pi=0$ GHz (\textbf{B} and \textbf{E}) and  near-resonant to the laser, $\Delta_j/2\pi=-0.3$ GHz (\textbf{C} and \textbf{F}). The vertical lines are at $\Delta_i/2\pi=0$ (black) and $\Delta_i/2\pi=-0.3$ GHz (orange).}
    \label{fig:detuning_cases}
\end{figure}

Figure~\ref{fig:detuning_cases} shows the measured intensity and second-order correlation at zero time delay $g^{(2)}(0)$ in transmission for different detunings under pulsed excitation. The detuning $\Delta_i$ between laser and QD$_i$ is swept for three different configurations: when the other QD, QD$_j$, is far detuned $\Delta_j\gg\Gamma_j$, on resonance $\Delta_j/2\pi=0$ GHz or near-resonant $\Delta_j/2\pi=-0.3$ GHz with the laser. We tested this for CW excitation before measuring the third-order correlation $g^{(3)}$. In this case, the strongest non-linear effects are observed at $\Delta_1/2\pi=\Delta_2/2\pi=-0.3$ GHz. Under pulsed excitation, the strongest non-linear interaction happens when $\Delta_1/2\pi=\Delta_2/2\pi=0$ GHz as can be seen in Fig.~\ref{fig:detuning_cases}. For the $g^{(3)}$ experiments we detune the QDs to be $\Delta_1/2\pi=\Delta_2/2\pi=-0.3$ GHz, however the magnitude of the detuning is almost identical to the spectral diffusion at $\sigma_{sd,1}/2\pi=0.3$ GHz and $\sigma_{sd,2}/2\pi=0.22$ GHz. Thus, the main text refers to the QDs as being resonant with the input light field. The impact of detuning and phase on $g^{(3)}$ is further discussed in section~\ref{sec:phase_scan} of the supplementary information. 

\subsection{Coupling up to three quantum dots}\label{sec:3QDs}
Among the four emitters, QD$_2$ is located in the first of the two electrically isolated waveguide sections (see section \ref{sec:qd_device}), while QD$_1$, QD$_3$, and QD$_4$ are embedded in the second section. This arrangement allows for individual electronic control of QD$_2$ through local electric Stark tuning~\cite{Chu.2023}. To control the emitters in the same waveguide section (QD$_1$, QD$_3$, and QD$_4$), an additional tuning mechanism is employed by applying an external magnetic field. The protocol for controlling triplets of emitters utilizes both tuning mechanisms. 

\begin{figure}[h!]
    \centering
    \includegraphics[width=1\linewidth]{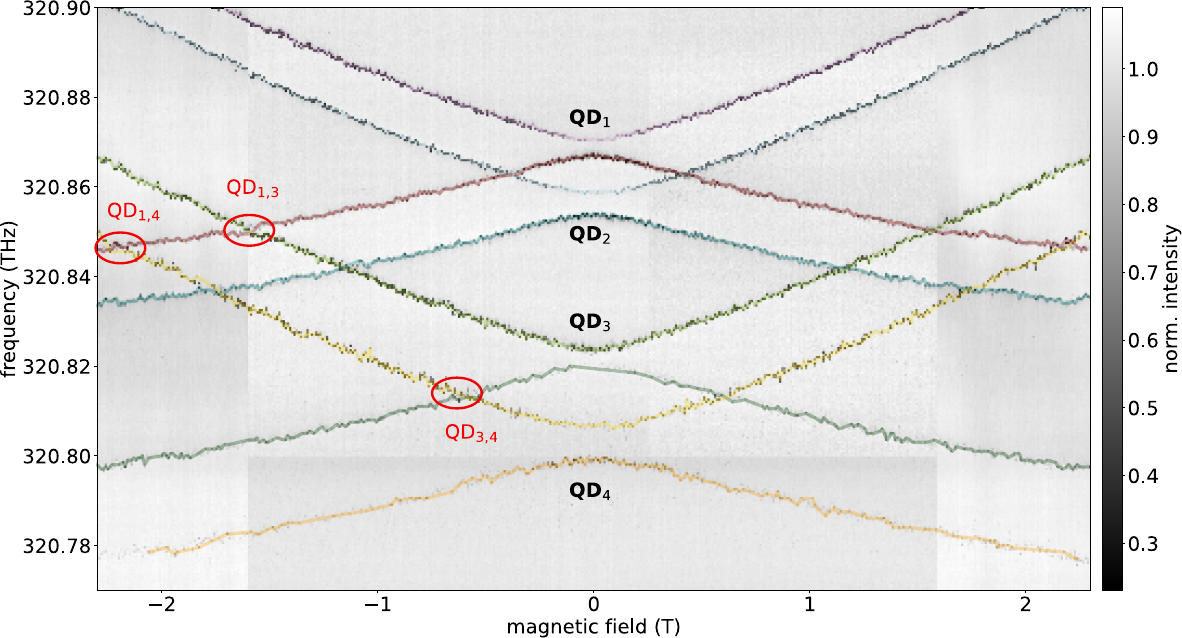}
    \caption{\textbf{Transmission spectra of four QDs.} Measurements of the transmission intensity as a function of the out-of-plane magnetic field $B_z$. The pairwise colored transition lines of the four QDs indicate the two dipoles of each emitter. The red ellipses indicate where pairs of QDs are tuned into resonance. In the measurement, a bias voltage $V_2=\SI{1.26}{V}$ is applied to QD$_2$ and the bias voltage applied to the second section, containing QD$_1$, QD$_3$, and QD$_4$, is $V_1=\SI{1.29}{V}$.}
    \label{fig:bbmap}
\end{figure}

First, pairs of emitters ($m=2$) are brought into resonance by applying a global magnetic field $B_z$ in a Faraday configuration perpendicular to the chip (see Ref.~\cite{Tiranov.2023}). The transmission spectrum as a function of the external magnetic field for four QDs is shown in Fig~\ref{fig:bbmap}. At $B_z=\SI{0}{T}$, the dipoles of each QD are shifted in energy due to fine-structure splitting. As the magnetic field strength increases, the dipoles exhibit opposite energy shifts because of the Zeeman effect. The resonant conditions are indicated by the intersections of the transition lines. To achieve resonant tuning of the low-frequency dipole of QD$_1$ (purple lines) and the high-frequency dipole of QD$_3$ (green lines), a magnetic field of $B_z=\SI{-1.56}{T}$ is applied. For tuning the low-frequency dipole of QD$_1$ into resonance with the high-frequency dipole of QD$_4$ (yellow lines), a magnetic field of $B_z=\SI{-2.17}{T}$ is used.

Next, to bring QD$_2$ into resonance with the pair QD$_{1,3}$(QD$_{1,4})$ a bias voltage of \SI{1.280}{V}(\SI{1.288}{V}) is applied to the first waveguide section. Therefore, the combination of electric and magnetic tuning enables control of up to three emitters for the triplets QD$_{1,2,3}$ and QD$_{1,2,4}$ ($m=3$). The pair QD$_{3,4}$ (resonant at $B_z=\SI{-0.56}{T}$) cannot be tuned into resonance with QD$_2$, as the frequency difference exceeds the electric tuning range of $\sim\SI{50}{GHz}$ (see Ref.~\cite{Henke.2026}). 

Additionally, pairs of emitters located in different sections of the waveguide are coupled through individual electric tuning. However, at $B_z=\SI{0}{T}$, QD$_4$ and QD$_2$ are spectrally too far from each other. Therefore, a magnetic field of $B_z=\SI{-2}{T}$ is applied to bring the low-frequency dipole of QD$_2$ and high-frequency dipole of QD$_4$ within mutual electric tuning range. 

Applying a magnetic field can impact the quality of a QD's light-matter interface in several ways. The magnetic field causes the transition frequencies of the QDs to become first-order sensitive to fluctuations in the magnetic field and to nuclear spin noise, which leads to increased spectral diffusion. In addition, the dipole moment polarization changes due to the magnetic field, which can reduce the coupling of the emitter's transition to the waveguide mode. Furthermore, at non-zero magnetic fields the transition frequency becomes susceptible to both spin dragging and anti-dragging effects on the QD dipoles~\cite{Hogele2012}. To mitigate these effects, the QDs are switched to the co-tunneling regime before each data acquisition, as described in Ref.~\cite{Antoniadis2022}. A switching frequency of \SI{100}{Hz} is used for the measurements. Note that higher switching frequencies can further improve the suppression of spin dragging and anti-dragging effects. Additionally, the sweeping of the laser frequency is randomized to minimize dragging or anti-dragging effects on the dipoles. 

Nevertheless, the effective light-matter interaction is reduced compared to the performance presented in Table~\ref{table:1}. The extracted parameters for each QD ($m=1$) at negative magnetic field ($B_z=\SI{-1.4}{T}$) are provided in Table~\ref{tab:qd parameter neg}. We note that these parameters can not be obtained fully independently~\cite{LeJeannic2021}. For comparison with $m=3$, the transmission behavior of the pairs QD$_{1,2}$ and QD$_{2,3}$ is also measured at $B_z=\SI{-1.4}{T}$. In future devices, negative effects associated with magnetic tuning can be avoided by employing only electrical tuning, which can be achieved by incorporating individual electrical tuning sections for each emitter.

The coupling phases between the emitters are fitted from lifetime measurements (described in Ref.~\cite{Tiranov.2023,Henke.2026}) to be $\phi_{1,2}=0.8\pi$, $\phi_{2,3}=1.2\pi$, and $\phi_{3,4}=\pi$. Note that the actual order of the emitters in the waveguide is QD$_{2}$, QD$_{3}$, QD$_{4}$, and QD$_{1}$ as opposed to QD$_{1}$, QD$_{2}$, QD$_{3}$, and QD$_{4}$ as depicted in the conceptual Fig.~\ref{fig:4}A in the main text. 

\begin{table}[h!]
    \centering
    \setlength{\tabcolsep}{12pt}
    \begin{tabular}{c|c|c|c|c}
     & QD$_1$ & QD$_2$&QD$_3$&QD$_4$   \\
     \hline\hline
         $\Gamma/ 2 \pi$, GHz &0.23& 0.21&0.17&0.16\\ \hline
    $\gamma_d / 2 \pi$, GHz&0.09&0.09&0.09&0.09\\ \hline
    $\beta$ & 0.44&0.37&0.45 &0.40  \\ \hline
    $\sigma_{sd} / 2\pi$, GHz & 0.25&0.25&0.25&0.25 \\ 
    \end{tabular}
    \caption{\textbf{Parameters of the four quantum dots.} Characterization of the four QDs parameters decay rate $\Gamma$, dephasing rate $\gamma_d$, $\beta$ factor, spectral diffusion $\sigma_{sd}$. The measurements for QD$_1$, QD$_2$, and QD$_3$ are taken at $B_z = \SI{-1.4}{T}$ and for QD$_4$ at $B_z=\SI{-2}{T}$. The coupling phases between the emitters are $\phi_{1,2}=0.8\pi$, $\phi_{2,3}=1.2\pi$, and $\phi_{3,4}=\pi$.}
    \label{tab:qd parameter neg}
\end{table}

\begin{figure}[h!]
    \centering
    \includegraphics[width=1\linewidth]{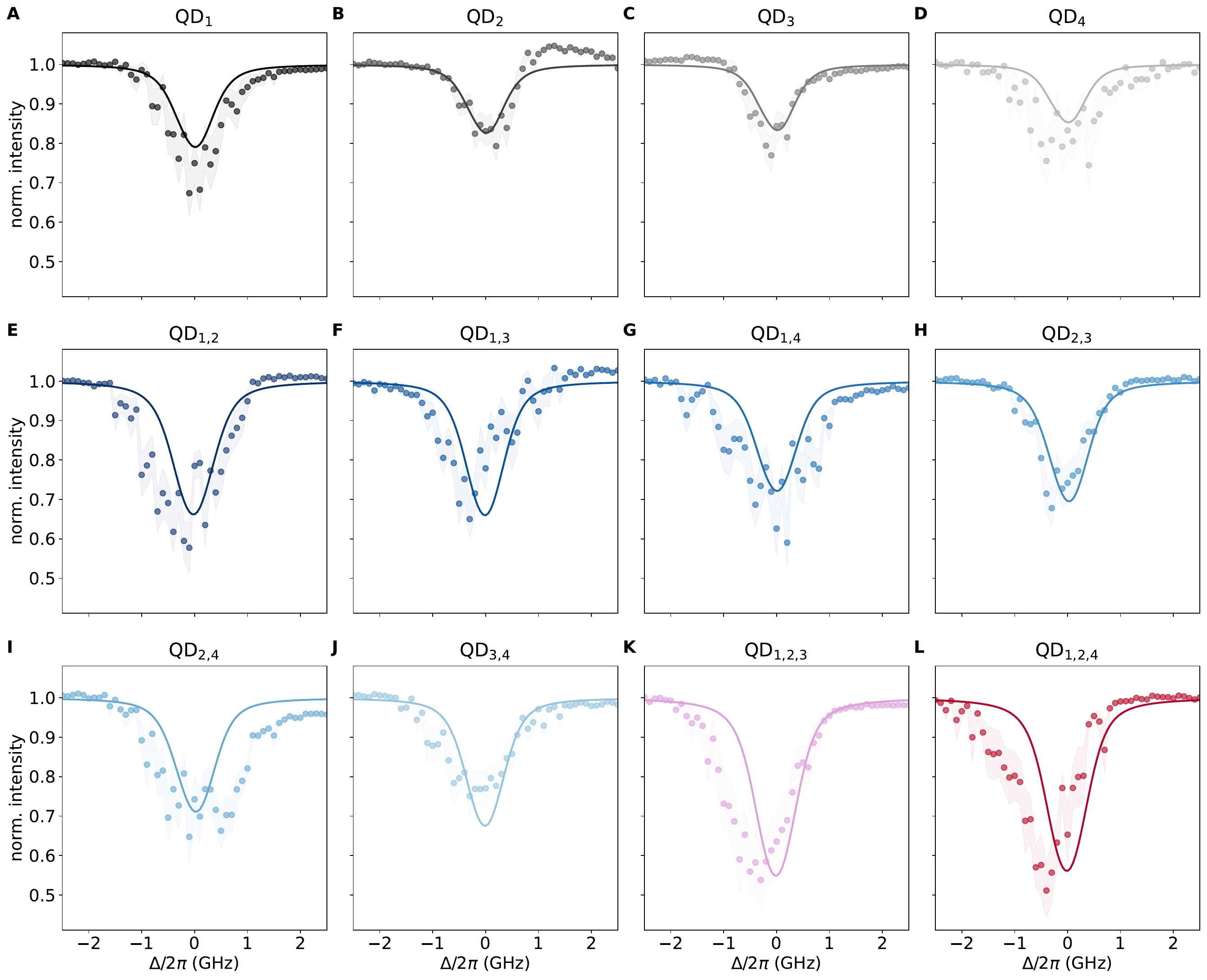}
    \caption{\textbf{Complete data set of coupling up to three emitters.} Measurements and simulations of transmitted intensity as a function of detuning for single emitters (gray lines), pairs of emitters (blue lines), and triplets of emitters (red lines). The data is normalized to the far off-resonant detuning and averaged over 15 runs. In the simulations, spectral diffusion is taken into account by sampling 1000 detuning values for each QD.}
    \label{fig:m3coupled}
\end{figure}

To investigate the collective effects of multiple emitters, transmission measurements for single emitters ($m=1$), pairs of emitters ($m=2$), and triplets of emitters ($m=3$) are compared (see Fig.~\ref{fig:4}C-E of the main text). The complete data sets and the corresponding simulations involving up to three emitters coupled to the waveguide are shown in Fig.~\ref{fig:m3coupled}. The parameters for the emitters used in the simulations are given in Table~\ref{tab:qd parameter neg}. In the case of one emitter ($m=1$), the depth of the transmission dip indicates the light-matter interaction with the incoming light field. When pairs of emitters are tuned resonant ($m=2$), collective effects of the emitter pair increase the depth of the transmission dip. This collective suppression is further enhanced for triplets of coupled emitters ($m=3$). These observations are in good agreement with the simulations. However, offsets in detuning and frequency shift might result from experimental calibration errors for zero detuning. Further, for collectively coupled emitters, the frequency shift becomes dependent on the coupling phase due to dispersive coupling effects~\cite{Henke.2026} and can indicate a collective Lamb shift in single-photon superradiance~\cite{Scully.2009}. Additionally, in contrast to the measurements, the theory model does not account for back reflections in the waveguide, which are responsible for the Fano profiles visible for individual QDs in Fig.~\ref{fig:4}C \cite{Javadi.2015, Thyrrestrup.2018}.

\subsection{Correlation measurements}\label{sec:complete_dataset}
The normalized \(n\)-th order photon correlation function is given by
\begin{equation}
    g^{(n)}(t_1, \ldots, t_n) = \frac{\langle a^\dagger (t_1) \ldots a^\dagger(t_n) a(t_n) \ldots a(t_1) \rangle}{ \langle a^\dagger(0) a(0) \rangle^n},
\end{equation}
where \(a^\dagger (t_i)\) and \(a(t_i)\) are the creation and annihilation operators at time $t_i$. The numerator of the above expression $G^{(n)}(t_1, ... t_n)$ denotes the unnormalized \(n\)-th order photon correlation function. Details about the normalization procedure are provided in section~\ref{sec:norm_avg}. 

The integration time for the forward scattering is \SI{152}{h}, and for the backwards scattering it is \SI{110}{h}. Every hour, the second QD was switched on or off to measure the scattering behavior for a single QD and both QDs semi-simultaneously for forward and backward scattering. The effective integration time is thus \SI{76}{h} and \SI{55}{h}, respectively. 
For the correlation measurements QD$_1$ and QD$_2$ are used exclusively.

\subsubsection{Third-order correlations} \label{sec:3rd_corr}
\begin{figure}[htp!]
    \centering
    \includegraphics[width=1.0\textwidth]{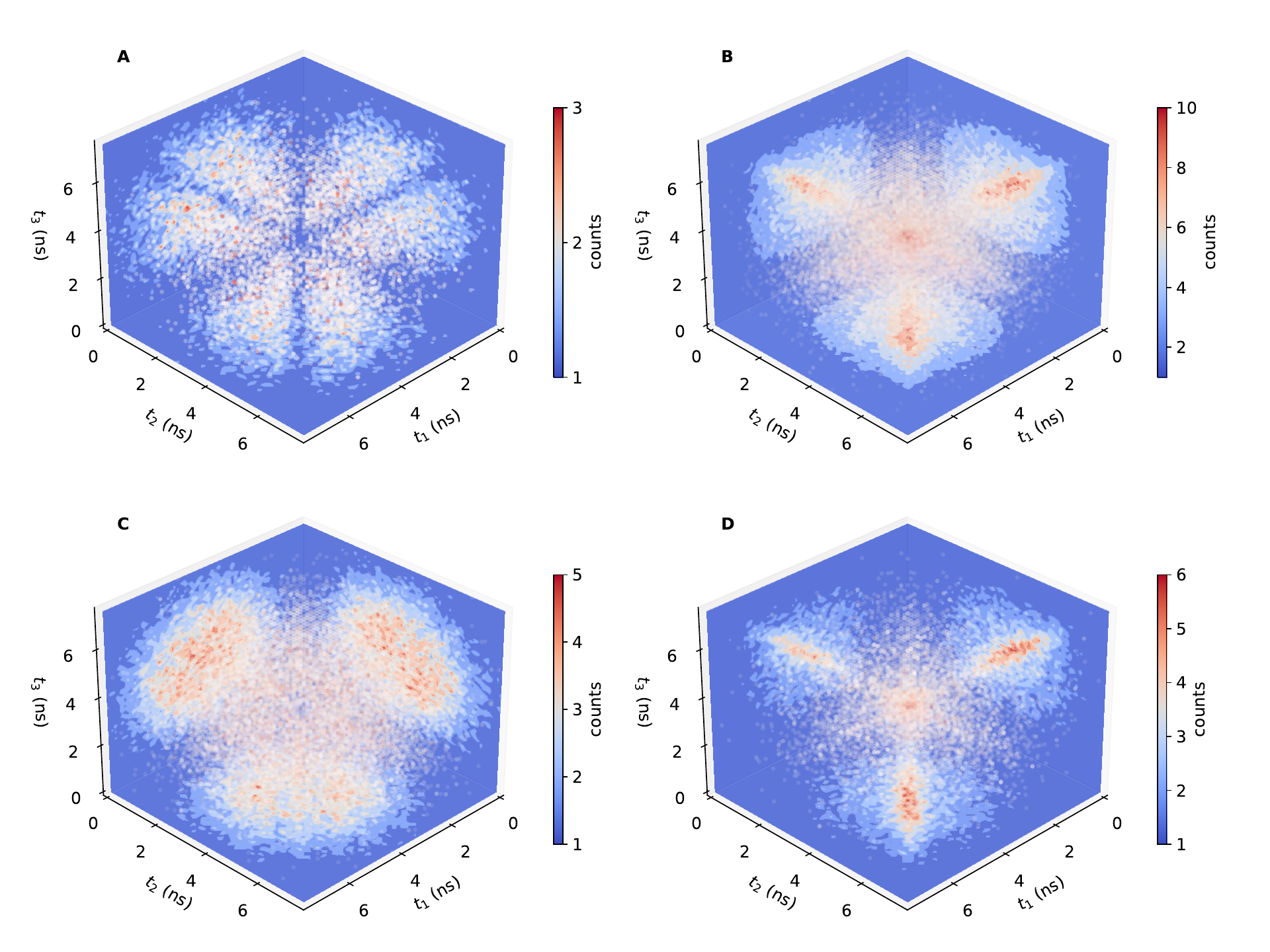}
    \caption{\textbf{Time-resolved third-order correlations.} Measurements of the time-resolved third-order correlation $G^{(3)}(t_1, t_2, t_3)$ of the backward-propagating field for (\textbf{A}) a single QD and (\textbf{C}) coupled QDs as well as for the forward-propagating field for (\textbf{B}) a single QD and (\textbf{D}) coupled QDs.} 
\label{fig:G3_3D}
\end{figure}

Figure~\ref{fig:G3_3D} shows 3D plots of the third-order correlation $G^{(3)}(t_1,t_2,t_3)$ for scattering in forward and backward direction off one or both QDs. Projections onto the faces are the two-photon correlations $G^{(2)}$  between two channels. The binsize is \SI{32}{ps}. 
In addition to three-photon correlations in which all photons originate from the same excitation pulse, correlations involving photons with time-delayed excitation are also measured.
In the partially correlated configuration, two photons are detected from the same excitation pulse, while the third photon is detected from the subsequent pulse, separated by one repetition period of \SI{20}{ns}. This configuration is described by the product \( G^{(2)}(t_i,t_j)G^{(1)}(t_k)\), as it combines a second-order correlation within a single pulse with an independent first-order correlation from the delayed pulse. In the uncorrelated configuration, each of the three detected photons is from a different excitation pulse. Since no photons share a common excitation event, the resulting statistics are fully independent and represent \(G^{(1)}(t_1)G^{(1)}(t_2)G^{(1)}(t_3)\).

\begin{figure}[htp!]
    \centering
    \includegraphics[width=1.0\textwidth]{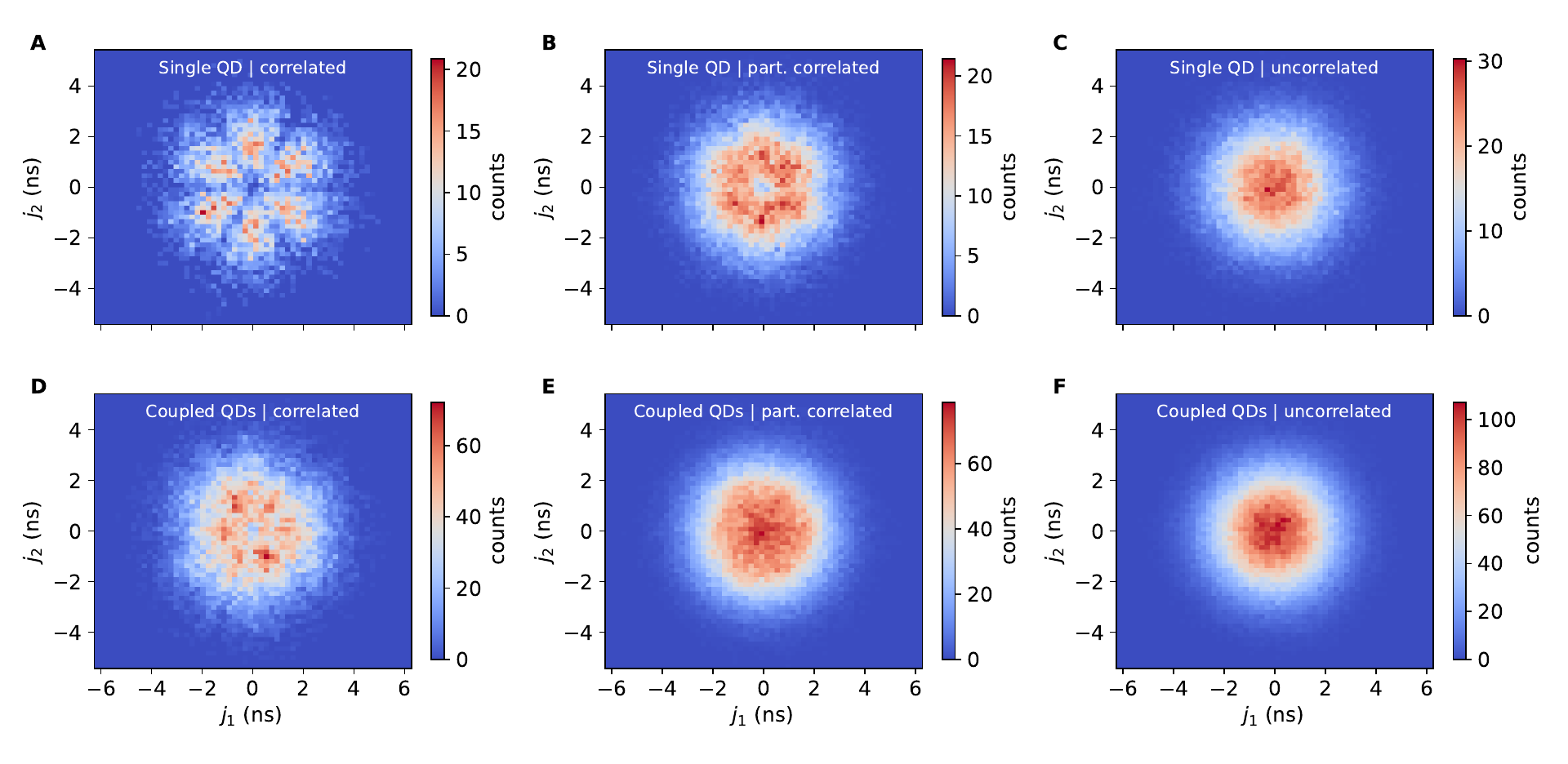}
    \caption{\textbf{Projected third-order correlation of backward-propagating field.} Measurements of the time-resolved third-order correlation $G^{(3)}(t_1, t_2, t_3)$ expressed in Jacobi coordinates $j_1$ and $j_2$ for a single QD (A, B and C) and coupled QDs (D, E and F). (\textbf{A} and \textbf{D}) For photons from the same excitation pulse, (\textbf{B} and \textbf{E}) two photons from the same pulse and one from a time-delayed pulse, and (\textbf{C} and \textbf{F}) uncorrelated measurement where all photons originate from different excitation pulses.} 
\label{fig:G3_b}
\end{figure}

\begin{figure}[htp!]
    \centering
    \includegraphics[width=1.0\textwidth]{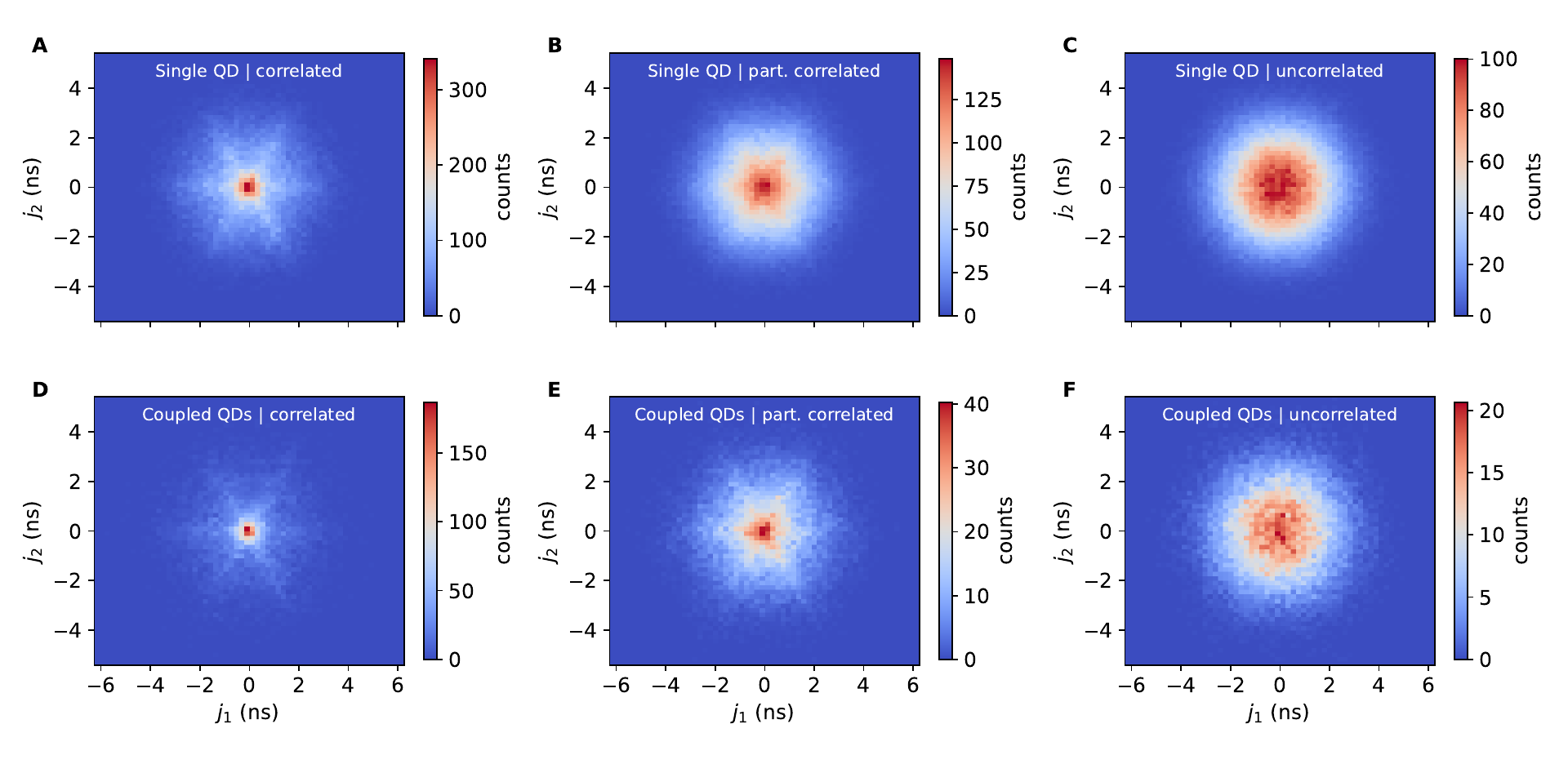}
    \caption{\textbf{Projected third-order correlation of forward-propagating field.} Measurements of the time-resolved third-order correlation $G^{(3)}(t_1, t_2, t_3)$ expressed in Jacobi coordinates $j_1$ and $j_2$ for a single QD (A, B and C) and coupled QDs (D, E and F). (\textbf{A} and \textbf{D}) For photons from the same excitation pulse, (\textbf{B} and \textbf{E}) two photons from the same pulse and one from a time-delayed pulse, and (\textbf{C} and \textbf{F}) uncorrelated measurement where all photons originate from different excitation pulses.} 
\label{fig:G3_f}
\end{figure}

 The third-order correlation measurements are transformed onto the Jacobi coordinates with $j_1 = (2t_1 - t_2 - t_3)/\sqrt{6}$ and $j_2 = (t_2-t_3)/\sqrt{2}$ and summed over the center-of-mass coordinate $j_0 = (t_1 + t_2 + t_3)/\sqrt{3}$.
The measured data are presented in Fig.~\ref{fig:G3_b} for backward scattering and in Fig.~\ref{fig:G3_f} for forward scattering.
The measurements of the partially and uncorrelated case are averaged over six possible channel delay combinations. Photon detection events are evaluated using time bins of \SI{128}{ps} width, corresponding to a rebinning with a factor 4 of the originally recorded data.
For the main text, the correlated measurements $G^{(3)}(j_1,j_2)$ are normalized by the respective uncorrelated measurement from three consecutive pulses, as it is related to the observable \(\langle a^\dagger(0) a(0) \rangle\) (see section~\ref{sec:norm_avg}). The normalized zero-delay values $g^{(3)}(0,0)$ are the ratio of the central areas for an integration area of \SI{417}{ps}$\times$\SI{362}{ps} corresponding to the four central bins from each histogram.

\subsubsection{Connected three-body correlations}\label{sec:3rd_connected}
We analyze the connected three-body correlation function utilizing the cumulant expansion~\cite{Kubo.1962, Stiesdal.2018, Plankensteiner.2022}:
\begin{equation}
   G^{(3)}_c(t_1,t_2,t_3) = 
   G^{(3)}(t_1,t_2,t_3) -\sum_{i<j,k} G^{(2)}(t_i,t_j)G^{(1)}(t_k) + 2G^{(1)}(t_1)G^{(1)}(t_2)G^{(1)}(t_3).
\end{equation} 
The contribution from the disconnected components is given by \(G^{(3)}_d = G^{(3)} - G^{(3)}_c\)~\cite{Drori.13.7.2023}. 
Experimentally, the connected three-body correlations are determined by recording the correlated $G^{(3)}(t_1,t_2,t_3)$, partially correlated coincidences from two consecutive pulses that is related to \( G^{(2)}(t_i,t_j)G^{(1)}(t_k)\), and uncorrelated coincidences from three consecutive pulses representing \(G^{(1)}(t_1)G^{(1)}(t_2)G^{(1)}(t_3)\). To compute $G^{(3)}_c$ and $G^{(3)}_d$ for the single QD and the coupled QDs in the forward or backward scattered fields, the plots in Fig.~\ref{fig:G3_b} and Fig.~\ref{fig:G3_f} are used. 

\subsubsection{Second-order correlations}\label{sec:2nd_corr}
\begin{figure}[htp!]
    \centering
    \includegraphics[width=0.9\textwidth]{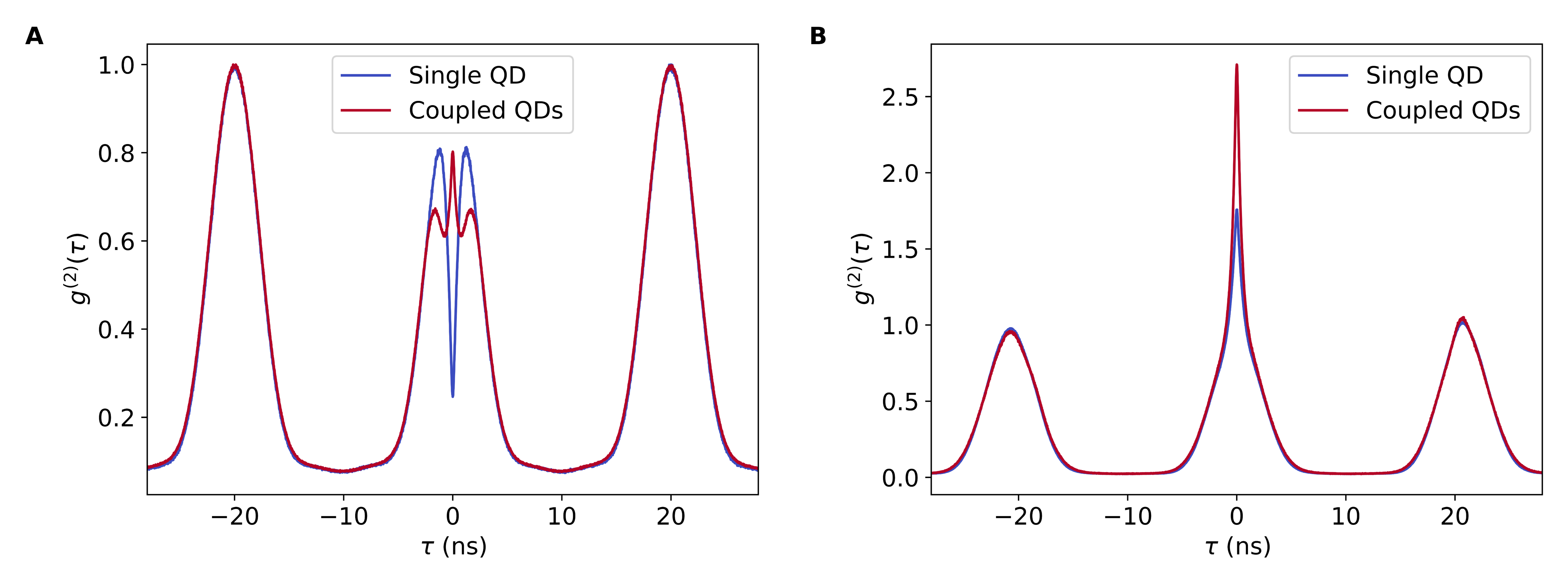}
    \caption{\textbf{Normalized second-order auto-correlation function.} (\textbf{A}) Measured normalized second-order auto-correlation function $g^{(2)}(0)$ for a single QD and coupled QDs of the backward-propagating field with $g^{(2)}(0)= 0.289(7)$ for the single QD and $g^{(2)}(0)= 0.790(2)$ for the coupled QDs and (\textbf{B}) of the forward-propagating field with $g^{(2)}(0)= 1.76(2)$ for the single QD and $g^{(2)}(0)= 2.71(8)$ for the coupled QDs.} 
\label{fig:s_g20}
\end{figure}

The normalized second-order correlations $g^{(2)}(\tau)$ for a single QD and coupled QDs are shown in Fig.~\ref{fig:s_g20} for the forward- and backward-propagating fields. The temporal resolution of these measurements is \SI{16}{ps}. The measurements are averaged over all three channel combinations and the normalization is performed for an integration window of \SI{256}{ps}. For the backward-propagating field (Fig.~\ref{fig:s_g20}B), anti-bunching is observed for a single QD ($g^{(2)}(0)= 0.289(7)$). This single-photon reflection dip is increased by a relatively low signal-to-noise ratio of ~$\approx20$, which originates from scattered laser light. For the coupled QDs, additionally, a bunched central peak ($g^{(2)}(0)= 0.790(2)$) is measured. This feature arises from two-photon reflection. In the forward direction (Fig.~\ref{fig:s_g20}B), an increased bunching is recorded for the coupled QDs ($g^{(2)}(0)= 2.71(8)$), as compared to the single QD ($g^{(2)}(0)= 1.76(2)$). We attribute this to the increased single-photon reflection of the coupled QDs, which suppresses the denominator of $g^{(2)}$. 

\begin{figure}[htp!]
    \centering
    \includegraphics[width=0.7\textwidth]{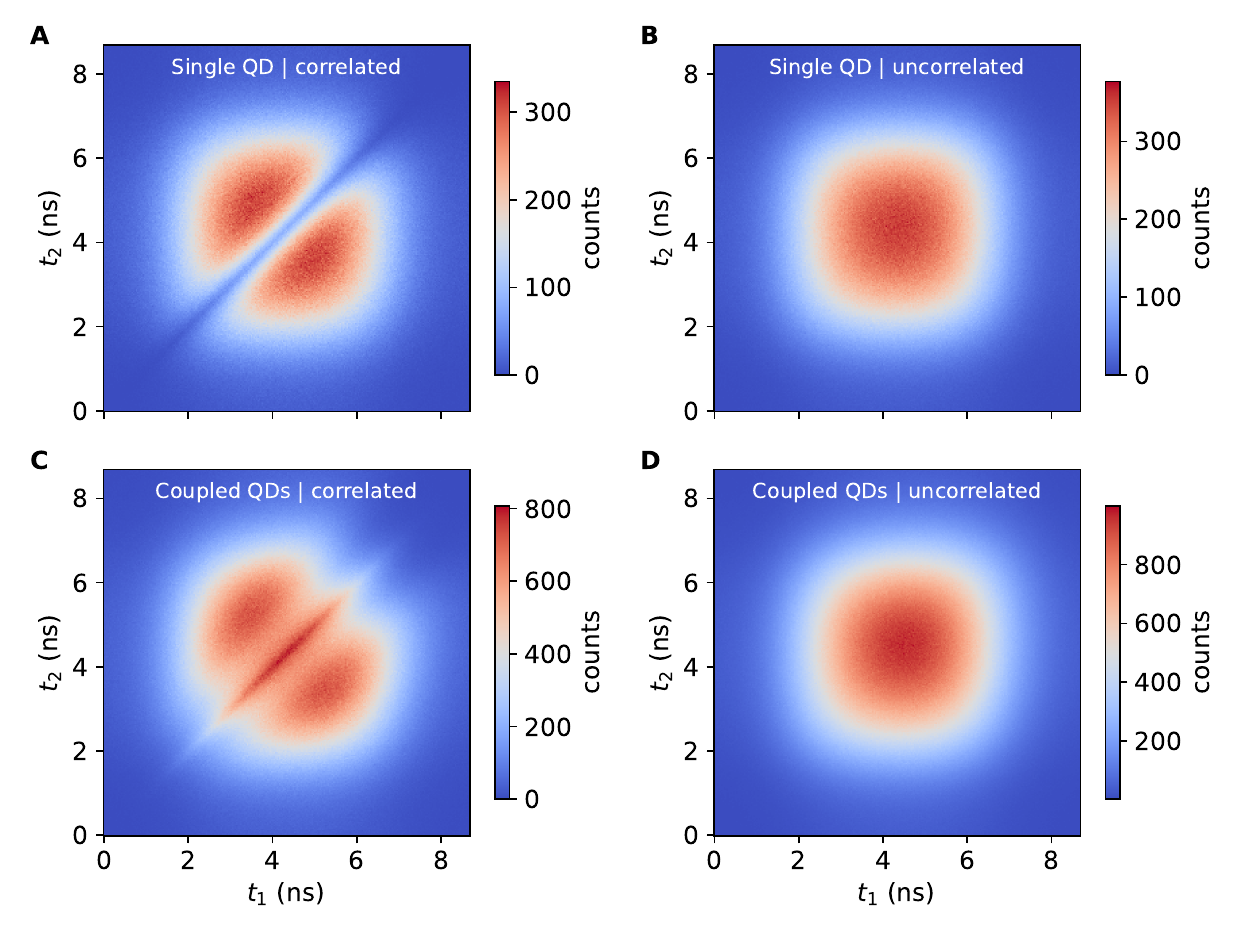}
    \caption{\textbf{Time-resolved second-order correlation of backward-propagating field.} (\textbf{A} and \textbf{C}) Measurements of the time-resolved second-order correlation $G^{(2)}(t_1, t_2)$ for a single QD and coupled QDs. (\textbf{B} and \textbf{D}) Uncorrelated measurement for a single QD and coupled QDs.} 
\label{fig:G2tt_b}
\end{figure}

\begin{figure}[htp!]
    \centering
    \includegraphics[width=0.7\textwidth]{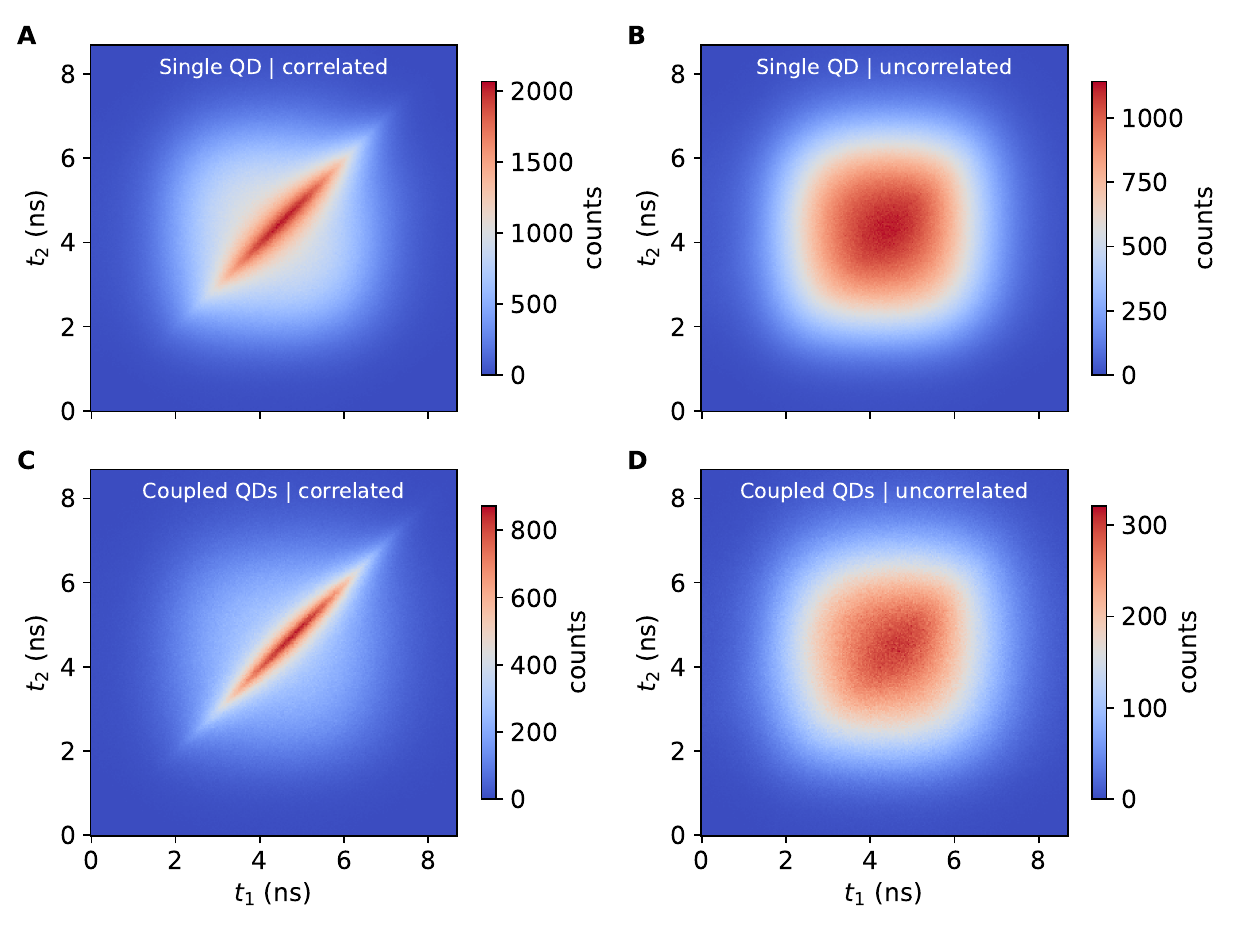}
    \caption{\textbf{Time-resolved second-order correlation of forward-propagating field.} (\textbf{A} and \textbf{C}) Measurements of the time-resolved second-order correlation $G^{(2)}(t_1, t_2)$ for a single QD and coupled QDs. (\textbf{B} and \textbf{D}) Uncorrelated measurement for a single QD and coupled QDs.} 
\label{fig:G2tt_f}
\end{figure}

The time-resolved second-order correlations $G^{(2)}(t_1,t_2)$ are shown for the backward- (Fig.~\ref{fig:G2tt_b}) and forward-propagating fields (Fig.~\ref{fig:G2tt_f}). The arrival of photons following the same excitation pulse (correlated) and from subsequent excitation pulses (uncorrelated) is recorded. The uncorrelated configuration effectively corresponds to \(G^{(1)}(t_1)G^{(1)}(t_2)\)  for scattering off a single QD and both QDs. The binsize is \SI{32}{ps} and the measurements are averaged over the three channel combinations. These measurements provide the data for the normalized plots shown in Fig.~\ref{fig:2} in the main text.

\clearpage
\subsection{Theoretical model}\label{sec:theory_model}
The theoretical model is described in Ref.~\cite{Sandberg.2023}. Most parameters follow Table~\ref{table:1} and section~\ref{sec:exp_setup}. However, the best agreement between the correlation experiments and simulations was achieved with $\gamma_{d,1}/2\pi = 0.09$ GHz, $\Delta_2/2\pi = -0.2$ GHz, and $\phi = 0.75\pi$, values within the experimental uncertainty.

\subsubsection{Simulation results}\label{sec:sim_results}
\begin{figure}[htp!]
    \centering
    \includegraphics[width=1.0\textwidth]{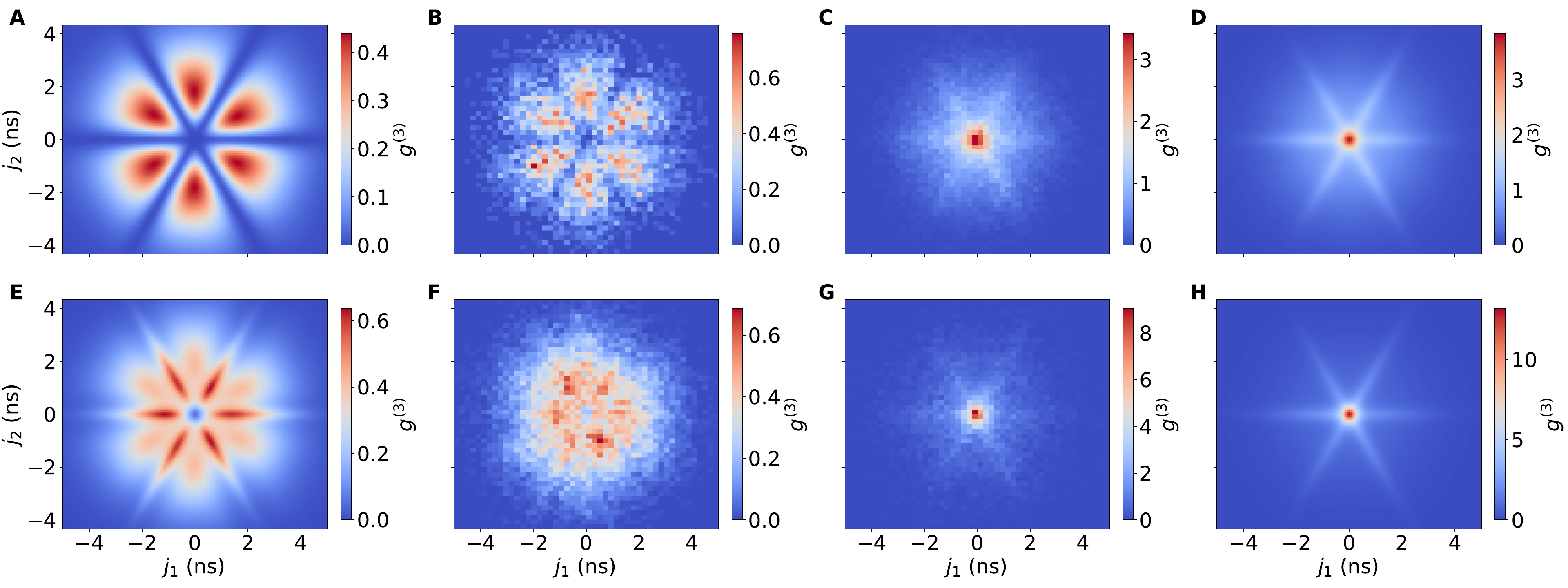}
    \caption{\textbf{Simulation and measurement of the normalized time-resolved third-order correlation, $g^{(3)}$.} (\textbf{A} and \textbf{B}) Simulation and measurement for backward-propagating field, and (\textbf{C} and \textbf{D}) forward-propagating field for a single QD. (\textbf{E} and \textbf{F}) Simulation and measurement for backward-propagating field, and (\textbf{G} and \textbf{H}) forward-propagating field for the coupled QDs.} 
\label{fig:S1}
\end{figure}

\begin{figure}[htp!]
    \centering
    \includegraphics[width=1.0\textwidth]{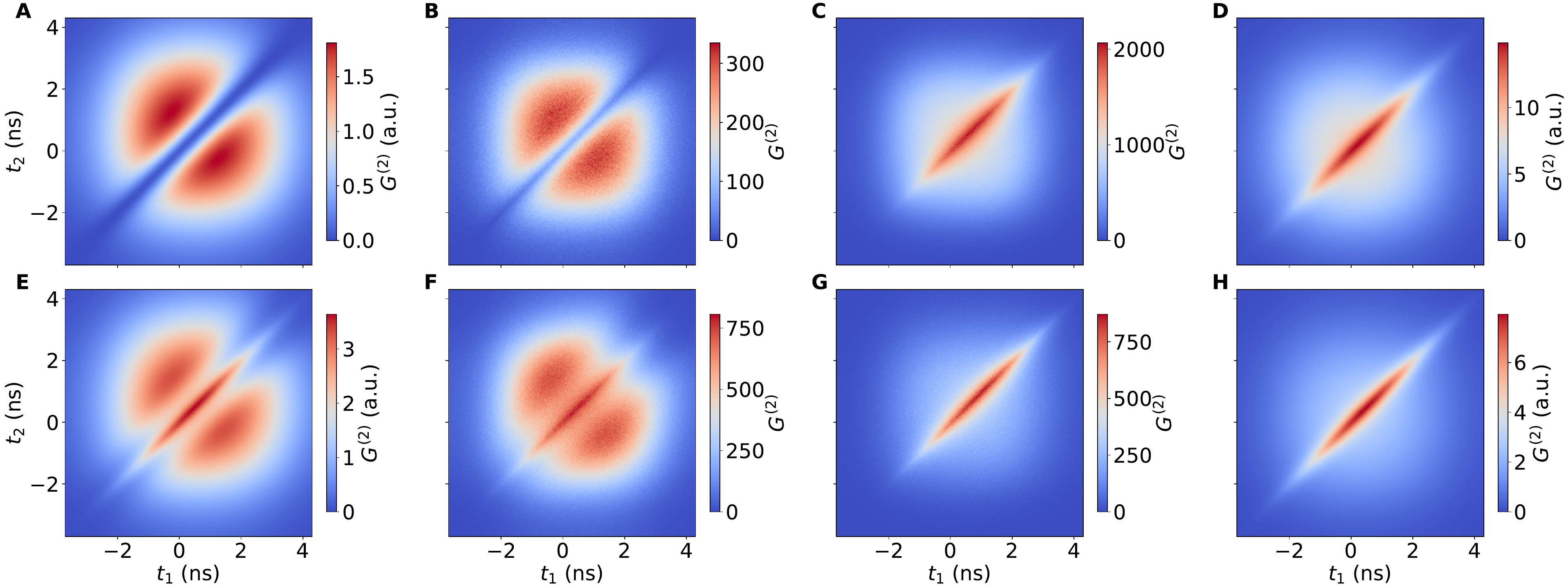}
    \caption{\textbf{Simulation and measurement of the time-resolved second-order correlation, $G^{(2)}(t_1,t_2)$.} (\textbf{A} and \textbf{B}) Simulation and measurement for backward-propagating field for one QD. (\textbf{C} and \textbf{D}) Measurement and simulation for forward-propagating field for one QD. (\textbf{E} and \textbf{F}) Simulation and measurement for backward-propagating field for two QDs. (\textbf{G} and \textbf{H}) Measurement and simulation for forward-propagating field for two QDs.} 
\label{fig:S2}
\end{figure}

\begin{figure}[htp!]
    \centering
    \includegraphics[width=1.0\textwidth]{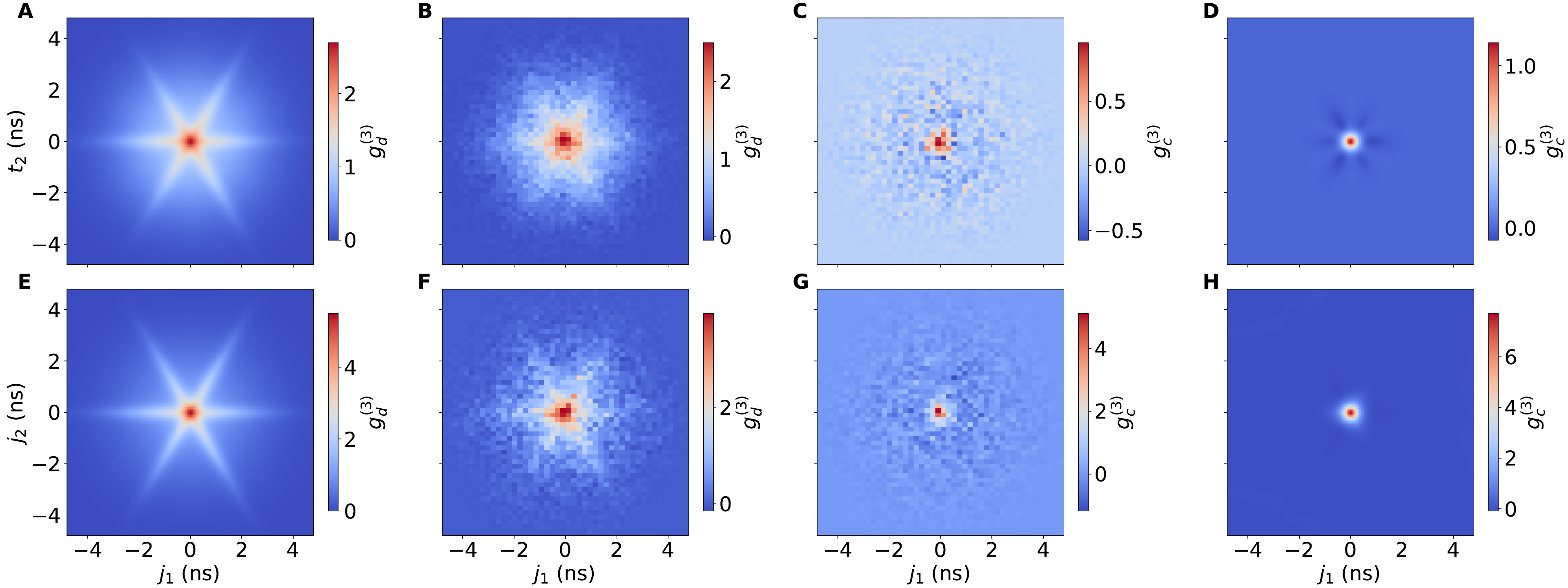}
    \caption{\textbf{Simulation and measurement of the connected three-body correlation, $g_c^{(3)}$.} (\textbf{A} and \textbf{B}) Simulation and measurement for backward-propagating field for one QD. (\textbf{C} and \textbf{D}) Measurement and simulation for forward-propagating field for one QD. (\textbf{E} and \textbf{F}) Simulation and measurement for backward-propagating field for two QDs. (\textbf{G} and \textbf{H}) Measurement and simulation for forward-propagating field for two QDs.} 
\label{fig:S3}
\end{figure}

Full scale images of the simulation results presented in the main text, shown next to the experimental data are given in Fig.~\ref{fig:S1} for $g^{(3)}$, in Fig.~\ref{fig:S2} for $G^{(2)}(t_1,t_2)$ and in Fig.~\ref{fig:S3} for $g_c^{(3)}$. In all cases we see a good agreement between data and simulation.

\subsubsection{Normalization and averaging} \label{sec:norm_avg}
In the experimental sequence, the correlated, partially correlated, and uncorrelated measurements (see section~\ref{sec:3rd_corr}, Fig.~\ref{fig:G3_b} and Fig.~\ref{fig:G3_f}) are obtained within one, two, and three consecutive pulses, respectively. The crucial point here is that spectral diffusion of the QDs occurs on a much longer time scale than the three consecutive pulses (\SI{60}{ns}). This means the intrinsic averaging of spectral diffusion in the experiment for the uncorrelated coincidences corresponds to  
\begin{equation}
    \overline{G^{(1)}(t_i) G^{(1)}(t_j) G^{(1)}(t_k)} = \sum_n^N p_n G^{(1)}_n(t_i) G^{(1)}_n(t_j) G^{(1)}_n(t_k)
    \label{eq:sd_avg1}
\end{equation}
where $p_n$ is the probability for certain detunings of the QDs given by the spectral diffusion, with $\sum_n p_n = 1$. This also applies to the partially correlated coincidences $\overline{G^{(2)}(t_i, t_j) G^{(1)}(t_k)}$ and uncorrelated two-photon coincidences $\overline{G^{(1)}(t_i) G^{(1)}(t_j)}$. 
Note, however, that the contributions of spectral diffusion can also be averaged differently, by e.g. first taking the average
\begin{equation}
    \overline{G^{(1)}(t_i)} \, \overline{G^{(1)}(t_j)} \, \overline{G^{(1)}(t_k)} = \sum_n^N p_n G^{(1)}_n(t_i) \sum_n^N p_n G^{(1)}_n(t_j) \sum_n^N p_n G^{(1)}_n(t_k). 
    \label{eq:sd_avg2}
\end{equation}
Experimentally, this can be achieved by simply taking the average of the photon number before calculating the uncorrelated coincidences or by counting uncorrelated coincidences of far-away pulses beyond the time-scale of spectral diffusion. Of course, also the partially correlated coincidences $\overline{G^{(2)}(t_i, t_j)} \, \overline{G^{(1)}(t_k)}$ and the uncorrelated two photon coincidences $\overline{G^{(1)}(t_i)}\, \overline{G^{(1)}(t_j)}$ can be averaged in this way. 

We want to emphasize that there are quantitative differences between the two averaging approaches and the results have slightly different interpretations. The first approach [Eq.~\eqref{eq:sd_avg1}] treats imperfections due to e.g., spectral diffusion or blinking equally for $G^{(n)}(t_1,...,t_n)$ and $G^{(1)}(t_1) ... G^{(1)}(t_n)$, corresponding to averaging the normalized correlation functions. This means that $G^{(n)}(t_1,...,t_n) / [ G^{(1)}(t_1) ... G^{(1)}(t_n) ]$ approaches one on time scales longer than the inverse decay times, but shorter than the time scale of e.g. spectral diffusion. 
On the other hand, the second approach [Eq.~\eqref{eq:sd_avg2}] averages coincidences, which corresponds to Glauber's definition of the normalized correlation function~\cite{Glauber1963}. If a quantitative characterization of the light source is made based on Glauber's definition of the correlation function, e.g.\ for non-classicality~\cite{Mandel1986} or quantum non-Gaussianity~\cite{Hotter2025}, one needs to average according to the second approach. For the shown theoretical simulations, we use the experimental (first) averaging approach for spectral diffusion, according to Eq.~\eqref{eq:sd_avg1}. Blinking is not taken into account in the simulations. For the parameters used in our simulations, we find that the second averaging approach would give a roughly 30 percent larger $g^{(3)}(j_1=0,j_2=0)$ for the transmission with a single QD and a roughly 50 percent larger with two QDs.

\subsubsection{Scaling of connected \(n\)th order correlations for many emitters}\label{sec:scaling}

\begin{figure}[ht]
    \centering
    \includegraphics[width=0.65\textwidth]{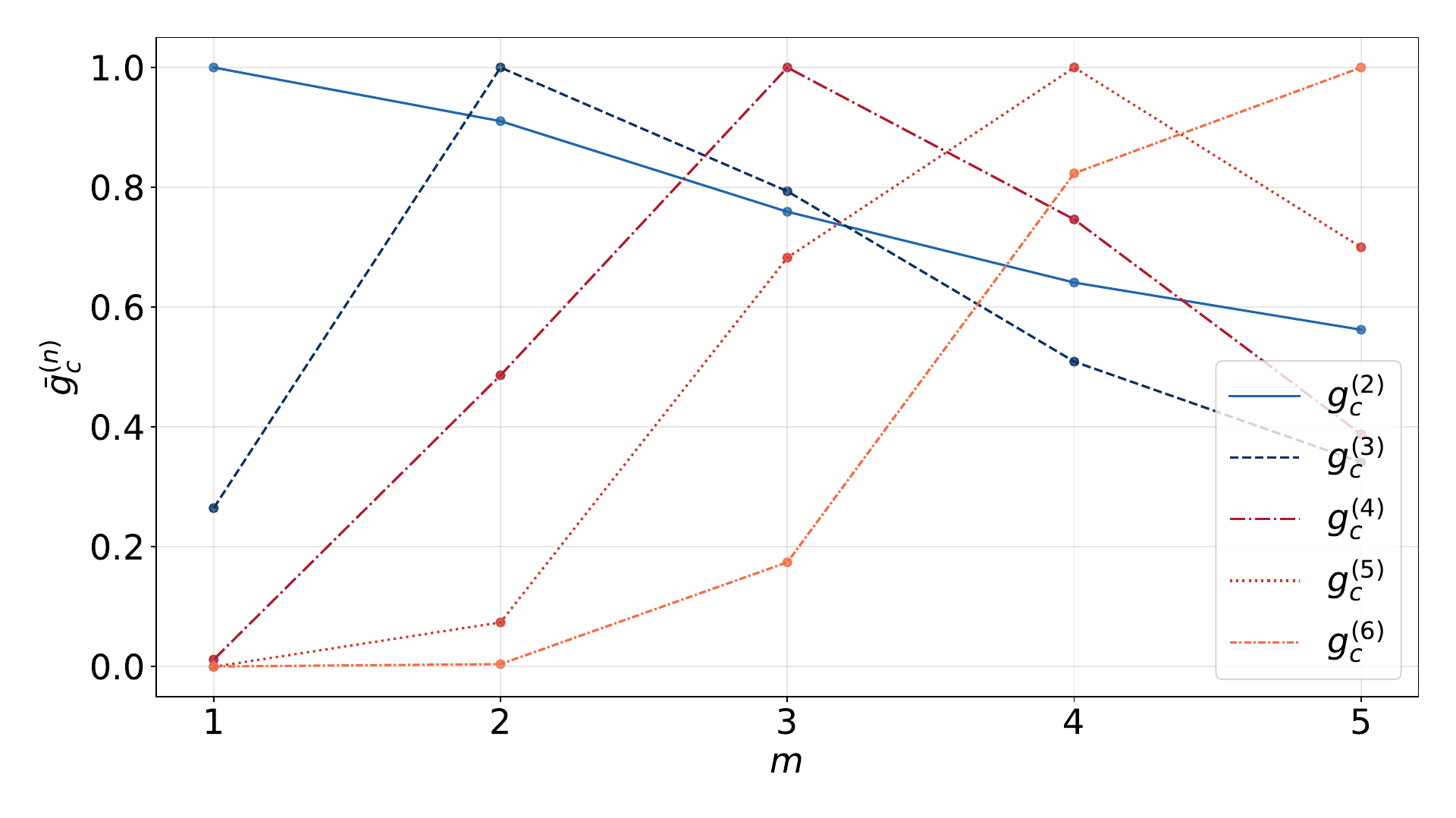}
    \caption{
    \textbf{Connected \(n\)th order correlations for \(m\) quantum emitters.} Scaling of the connected \(n\)th order correlation \(g^{(n)}_c\) as a function of the number of coupled emitters $m$. For clarity, the plot shows the normalized connected component \(\bar{g}^{(n)}_c\). The parameters for the simulations are \(\phi=0\), \(\beta = 0.95\), \(\gamma_d/2\pi=0.09\) and \(\sigma = 1\, \text{ns}\). 
    }
    \label{fig:S15}
\end{figure}

As explained in the main text, the nonlinearity of the system is based on the intuitive picture that $m$ QDs are able to reflect $m$ photons, but additional photons can stimulate the emission into the forward direction. 
This means that the transmission of $m$ photons is suppressed, which advances the $(m+1)$-th order joint cumulant $g^{(n)}_c$. This can be seen in Fig.~\ref{fig:S15}, where the $n$-th order joint cumulant for different numbers of emitters $m$, peaks at $n=m+1$. We normalized here $g^{(n)}_c$ to the peak value for each $n$ because the actual values for larger $n$ are very large. We want to note here that an appropriate pulse width $\sigma$ is required to observe the described effect. For too short pulses, only a small part of the wavepacket is resonant with the transition, whereas for too long pulses, the nonlinearity is reduced. Furthermore, one should be aware that the effective collective decay rate of the $m$ QDs system is also superradiantly enhanced. This means that for more QDs, a shorter pulse is required to reach the nonlinear regime. 

\subsubsection{Phase scan}\label{sec:phase_scan}
\begin{figure}[ht]
    \centering
    \includegraphics[width=0.55\textwidth]{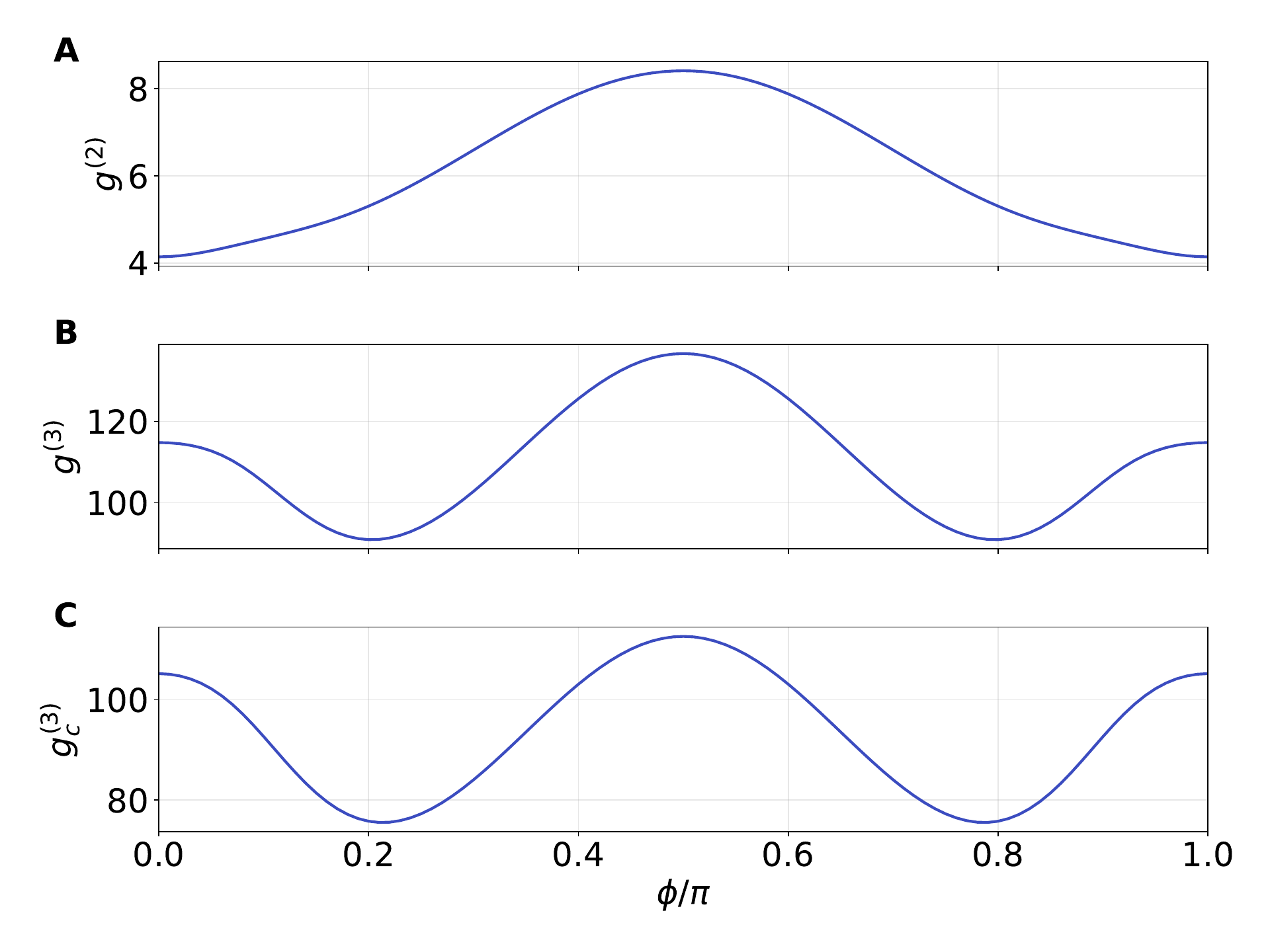}
    \caption{\textbf{Coupling phase-dependent collective quantum nonlinear effects.} The simulations of (\textbf{A}) \(g^{(2)}\), (\textbf{B}) \(g^{(3)}\), and (\textbf{C}) \(g^{(3)}_c\) show the nonlinear behavior as a function of the coupling phase \(\phi\) between the QDs. The parameters for the simulations are \(\beta = 0.95\), a dephasing rate \(\gamma_d/2\pi=0.09\), an input pulse width \(\sigma = 1\, \text{ns}\), and exclude spectral diffusion.
    }
    \label{fig:phase}
\end{figure}

In an ideal experimental setup, where the QDs do not experience spectral diffusion, we expect a strong nonlinearity for resonant light ($\Delta_1 = \Delta_2 = 0$) at a phase between the QDs of $n \cdot \pi$ with $n \in \mathbb{N}$, which corresponds to fully dissipative coupling, see Fig.~\ref{fig:phase}A-C. Besides assuming no spectral diffusion we use similar parameters as in the experiment for the simulation, i.e.\ $\Gamma / 2\pi =  0.388 \, \mathrm{GHz}$, $\beta = 0.95$, $\gamma_\mathrm{d} / 2\pi = 0.09 \, \mathrm{GHz}$ and a (shorter) Gaussian input pulse with width $\sigma = 1 \, \mathrm{ns}$. All parameters are the same for both QDs. We can see that $g^{(2)}$ has a minimum and $g^{(3)}$ a maximum for $\phi = n \cdot \pi$, which results in a maximum for $g_c^{(3)}$. 
Although $g^{(2)}$ has a maximum at $\phi = \pi/2$ there is still a maximum also for $g_c^{(3)}$ due to the larger value of $g^{(3)}$.  

Overall, we see that if spectral diffusion can be reduced in the device (and shorter pulses are used), a significantly larger connected component $g_c^{(3)}$ can be achieved, especially by working on resonance. In this case, a large connected component can be obtained with any phase $\phi$ between the QDs. 

\end{document}